\documentclass[fleqn,12pt,numbers=noenddot]{wlscirep}
\usepackage[utf8]{inputenc}
\usepackage[T1]{fontenc}
\usepackage[usestackEOL]{stackengine}

\usepackage{url} %
\usepackage{lineno}
\usepackage{multibib}
\newcites{M}{References}
\newcites{S}{References}
\newcites{L}{References}

\let\oldequation\equation
\let\oldendequation\endequation

\renewenvironment{equation}
  {\linenomathNonumbers\oldequation}
  {\oldendequation\endlinenomath}
  
\usepackage{multirow} %

\title{
Building Wet Planets through High-Pressure Magma-Hydrogen Reactions}

\author[1,4,*]{H. W.~Horn}
\author[2]{A.~Vazan}
\author[3]{S.~Chariton}
\author[3]{V. B. Prakapenka}
\author[1,*]{S.-H. Shim}
\affil[1]{School of Earth and Space Exploration, Arizona State University, Tempe, Arizona, USA.}
\affil[2]{Astrophysics Research Center, Department of Natural Sciences, Open University of Israel, Ra'anana, Israel.}
\affil[3]{Center for Advanced Radiation Sources, University of Chicago, Chicago, Illinois, USA.}
\affil[4]{Now at Lawrence Livermore National Laboratory, USA.}
\affil[*]{hallensu@asu.edu; sshim5@asu.edu}

\makeatletter
\renewcommand{\@maketitle}{%
{%
\thispagestyle{empty}%
\vskip-36pt%
{\raggedright\sffamily\bfseries\fontsize{20}{25}\selectfont \@title\par}%
\vskip10pt
{\raggedright\sffamily\fontsize{12}{16}\selectfont  \@author\par}
\vskip25pt%
}%
}%
\makeatother

\begin{document}

\maketitle

\textbf{Close-in transiting sub-Neptunes are abundant in our galaxy \cite{fulton2017california}. 
Planetary interior models based on their observed radius-mass relationship suggest that sub-Neptunes contain a discernible amount of either hydrogen (dry planets) or water (wet planets) blanketing a core composed of rocks and metal \cite{bean2021nature}. 
Water-rich sub-Neptunes have been believed to form farther from the star and then migrate inward to their present orbits \cite{bitsch2021Dry}. 
Here, we report experimental evidence of reactions between warm dense hydrogen fluid and silicate melt that releases silicon from the magma to form alloys and hydrides at high pressures.
We found that oxygen liberated from the silicate melt reacts with hydrogen, producing a significant amount of water up to a few tens of weight percent, which is much greater than previously predicted based on low-pressure ideal gas extrapolation \cite{misener2023Atmospheresa,schlichting2022Chemical}. 
Consequently, these reactions can generate a spectrum of water contents in hydrogen-rich planets, with the potential to reach water-rich compositions for some sub-Neptunes, implying an evolutionary relationship between hydrogen-rich and water-rich planets.
Therefore, detection of a large amount of water in exoplanet atmospheres may not be the optimal evidence for planet migration in the protoplanetary disk, calling into question the assumed link between composition and planet formation location.
}

\thispagestyle{empty}

Water is an important building block of planets and a key ingredient for their potential habitability.
The notion that water is incorporated into planetary bodies through condensation at sufficiently low temperatures in the outer proto-planetary disk seems to explain the architecture of the solar system well: the inner planets are mostly dry and rocky sometimes with a small amount of water delivered from the outer solar system while Uranus and Neptune---which are believed to be ice giants---exist outside of the snow line \cite{morbidelli2000Source}.
However, some nebular hydrogen-rich gas was possibly captured and stored as H$_2$O in the Earth's deep interior after reacting with silicates \cite{ikoma2006constraints,hallis2015evidence,young2023Earth}.

NASA's \emph{Kepler} mission has discovered many of close-in transiting exoplanets between Earth-like and Neptune-like in size (1--4~Earth radius, $R_{E}$) and density\cite{howard2012Planet}.
Planets with radii of 2--4~$R_E$ are typically modeled with rocky cores and envelopes dominated by either hydrogen or water, making their composition and formation ambiguous (where ``core'' refers to the non-volatile interior, as commonly defined in exoplanet literature).
A drop in relative frequency of occurrence among such planets was identified at 1.5--2\,$R_{E}$ (i.e., the radius valley) \cite{fulton2017california}, which appears to divide planets into two groups: rocky + metallic Earth-like density with tiny or no atmosphere at 1--1.5\,$R_{E}$ (i.e., super-Earths), and rocky + metallic core with thick atmosphere at 2--4\,$R_{E}$ (i.e., sub-Neptunes).
Models involving massive gas loss can explain the radius valley well \cite{owen2013kepler,ginzburg2018Corepowered}, which seems to support hydrogen-dominated atmospheres, while other models explain the radius valley by two separate populations of dry and water-rich planets \cite{zeng2019growth,venturini2020jupiter}.
Moreover, some systems may indeed host water-rich sub-Neptunes \cite{luque2022Density,piaulet2022Evidence,piaulet-ghorayeb2024JWST}. 
An important question for water-rich sub-Neptune models is how can planets with a significant water mass fraction exist in close-in orbits if water mainly condensed and was incorporated into planets in the outer part of the systems during their formation \cite{bitsch2021Dry}. 

In classical models of sub-Neptunes with hydrogen-rich envelopes, hydrogen is typically assumed not to react with silicates and metals in the core, though physical mixing with silicate magma has been considered \cite{hirschmann2012solubility,kite2019superabundance}. 
At low pressures and high temperatures, hydrogen can reduce cations in silicates and oxides to metals \cite{sabat2014ReductionOxideMinerals}, releasing oxygen from melts and producing water through hydrogen oxidation. 
Experimental data show that hydrogen can reduce Fe$^{2+}$ to Fe$^{0}$ in molten silicates, generating water at low pressure\cite{ikoma2006constraints,kimura2022Predicted,krissansen-totton2024Erosion}, and this trend extends to higher pressures \cite{horn2023reaction}. 
However, in protoplanetary disks, most iron is metallic, with only limited Fe$^{2+}$ and Fe$^{3+}$ available.  
Recent experiments show that Mg remains oxidized in dense hydrogen, but small amounts of MgH$_2$ can form, releasing oxygen for water production \cite{kim2023Stability}.  
Thermodynamic models based on low-pressure data suggest that hydrogen cannot reduce a significant amount of Si$^{4+}$ in silicates to produce water\cite{ikoma2006constraints,schlichting2022Chemical}. 
A modeling study indicates that enhanced Si partitioning into metal under reducing conditions could yield Earth-like water abundances (${\sim}10^{-2}$~wt\%) \cite{young2023Earth}. 
However, experiments show that SiO$_2$ can dissolve in hydrogen at 2--15~GPa and 1400--1700~K, with Si--O bonds breaking to form Si--H, suggesting that pressure may promote hydrogen-silicate reactions \cite{shinozaki2013Influence,shinozaki2014formation}.

If the mass-radius of sub-Neptunes is modeled with a rock + metal core with a hydrogen-dominated envelope, the pressure at the core-envelope boundary (CEB) can exceed a few GPa \cite{stokl2016Dynamical}.
This blanketing envelope can also decrease heat loss such that silicates and metals of the core remain molten for billions of years \cite{vazan2018Contribution}.
Under these conditions, hydrogen is a dense fluid.
Therefore, it is essential to understand the behavior of the hydrogen-silicate system at the high pressures and temperatures $(P\text{-}T)$ that exist at such a CEB\@.
However, to our knowledge, such data does not exist.
In this study, we combined pulsed laser heating\cite{goncharov2010x} with the diamond-anvil cell to mitigate diamond embrittlement which has been a major problem in melting silicate and metal in a pure hydrogen medium to acquire key data for understanding possible hydrogen-silicate reactions at the $P\text{-}T$ conditions expected at the CEB in sub-Neptunes (Fig.~\ref{fig:fib}; Extended Data Figs~\ref{fig:SI:PT} and \ref{fig:SI:amb-PT}; Extended Data Table~\ref{tab:SI:run-table}; Supplementary Discussion~\ref{SI:section:diamond-embrittlement}).

\section*{High-Pressure Magma-Hydrogen Reactions}

A starting mixture of San Carlos olivine, $\mathrm{(Mg_{0.9}Fe_{0.1})_{2}SiO_{4}}$, 
and iron metal was melted in a hydrogen medium at 8~GPa (Extended Data Table~\ref{tab:SI:run-table}, run~SCO-3).
After heating, high-$P$ X-ray diffraction (XRD) patterns showed no peaks from silicates, including olivine (Fig.~\ref{fig:xrd}a). 
Instead, diffraction lines of B2-$\mathrm{Fe}_{1-y}\mathrm{Si}_y$ were observed, showing that Si$^{4+}$ in silicate melt was reduced to Si$^{0}$ and alloyed with Fe metal. 
After heating, Mg remains as MgO periclase.
The unit-cell volume of MgO after decompression to 1~bar indicates no detectable Fe$^{2+}$ in the phase. Therefore, all Fe$^{2+}$ originally in the silicate is also reduced to Fe$^0$ metal, implying the following reaction: %
\begin{equation}
\mathrm{(Mg_{0.9}Fe_{0.1})_{2}SiO_{4}} + 0.8\mathrm{Fe} + 2.2\mathrm{H}_2 \rightarrow 
\mathrm{FeSi} + 
1.8\mathrm{MgO} + 2.2\mathrm{H_{2}O}.\label{eq:ol-Si-reduction}
\end{equation}

In our experiments, 20~wt\% of Fe metal was mixed with olivine and a smaller amount of Fe metal is formed from the reduction of Fe$^{2+}$ in olivine.
If all the Fe metal is consumed to form FeSi (i.e., $y=0.5$ in Fe$_{1-y}$Si$_y$), a maximum of 86\% of the Si$^{4+}$ in olivine can be reduced (Supplementary Code~1).   
However, the measured unit-cell volume of the Fe-Si alloy indicates $y = 0.27$ in $\mathrm{Fe}_{1-y}\mathrm{Si}_y$ (Supplementary Discussion~\ref{SI:section:olivine-composition}--\ref{SI:section:fay-spifield}).
Therefore, only 32\% of Si$^{4+}$ in olivine was reduced to form alloy with Fe metal.  
Furthermore, FeH$_x$ alloys in the fcc and dhcp structures ($x=0.22{-}0.4$) were found after heating, thus decreasing the amount of Fe metal available to form Fe-Si alloys. 
The observation of complete disappearance of silicates after melting in dense hydrogen fluid requires an additional process to remove Si$^{4+}$ from silicates.

\begin{figure}[!ht]
\centering
\includegraphics[width=\textwidth]{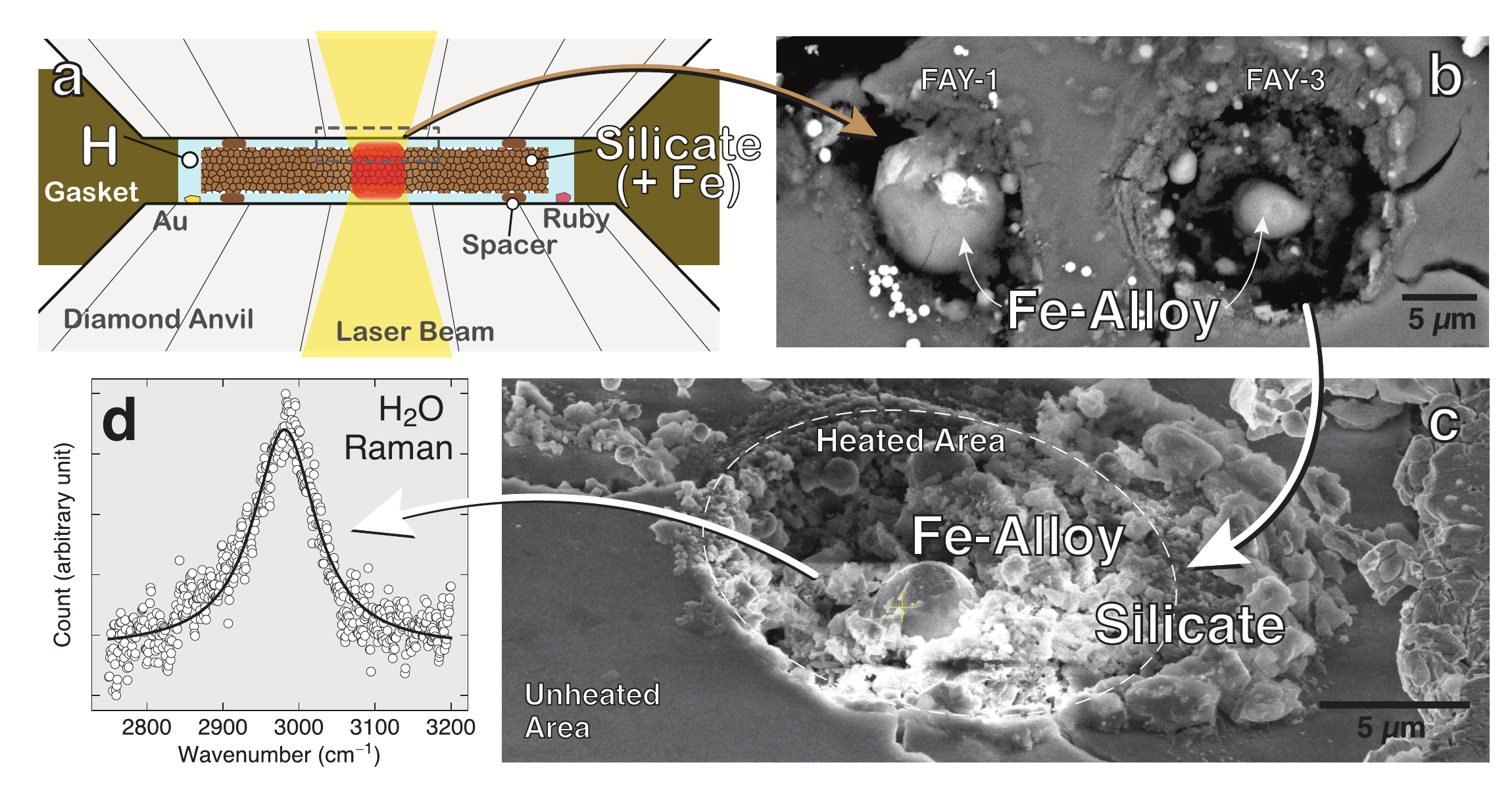}
\caption{{\bf Laser-heated diamond-anvil cell (LHDAC) experiments on silicate melts in a hydrogen medium.} 
{\bf a,} Schematic diagram for the experimental setup. 
The spacers (small single grains of the same material from which the foil was made) separate the sample foil from diamond anvils allow hydrogen to surround the sample.
During laser heating (the red area at the center) of the silicate sample, hydrogen penetrated into the grain boundaries of the sample foil and immediately above and below were heated by thermal conduction.
{\bf b,} An SEM image of two laser heated areas of a fayalite sample.
FAY-1 was heated at 6~GPa and 3017~K and FAY-3 was heated at 11~GPa and 2898~K\@.
The spheres at the center of the heated areas are Fe-rich alloys formed by hydrogen-silicate reaction. %
{\bf c,} An SEM image of FAY-3 from an angle for a wider area.
{\bf d,} Raman-active OH vibration from H$_2$O ice after heating silica + Fe metal in a H$_2$ medium at 14~GPa. A full 2D Raman map of the heated area is included in Extended Data Fig.~\ref{fig:SI:map}c.}\label{fig:fib}
\end{figure}

Previous experiments at similar pressures but lower temperatures\cite{shinozaki2013Influence,shinozaki2014formation} showed the release of Si$^{4+}$ from silicate and dissolution in dense fluid hydrogen as SiH$_4$: 
\begin{equation}
\mathrm{Mg_{2}SiO_{4}} + 4\mathrm{H}_2 \rightarrow 
\mathrm{SiH_4} + 2\mathrm{MgO} + 2\mathrm{H_{2}O}.\label{eq:ol-Si-dissolve}
\end{equation}
We detected the same Si--H bond vibration in melted areas in our experiment (Extended Data Fig.~\ref{fig:SI:SiH4} and Supplementary Discussion~\ref{SI:section:fay-50H}). 
Both the reduction and hydride formation reactions (reactions~\ref{eq:ol-Si-reduction} and \ref{eq:ol-Si-dissolve}, respectively) involve the release of O which can then react with H forming H$_2$O.
In fact, the OH vibration from H$_2$O was unambiguously detected in melted areas with Raman spectroscopy (Fig.~\ref{fig:fib}d; Extended Data Fig.~\ref{fig:SI:map}).
The same behavior was also observed at pressures up to 22~GPa in multiples of experimental runs (Supplementary Discussion~\ref{SI:section:olivine-composition}-\ref{SI:section:ringwoodite-synthesis} and Extended Data Fig.~\ref{fig:SI:xrd_olv21})

When olivine + Fe was heated at 42~GPa to temperatures below melting \cite{shen1995melting}, olivine transformed to bridgmanite (bdm) and ferropericlase (fp) (Fig.~\ref{fig:xrd}c). 
We further heated the synthesized bdm + fp above the melting temperature of bdm.
The majority of bdm dissociates upon melting in a hydrogen medium (Fig.~\ref{fig:xrd}d).
The SiO$_2$ component reacts with H and Fe to form B2~$\mathrm{Fe}_{1-y}\mathrm{Si}_y$ (Extended Data Fig.~\ref{fig:SI:map}a,b).
The volume measured for quenched fp suggests the 
complete reduction of Fe$^{2+}$ to Fe$^0$ (Supplementary Discussion~\ref{SI:section:bridgmanite-recovered-composition}). 

With fayalite (Fe$_2$SiO$_4$) and SiO$_2$ + Fe metal, we observed consistent behaviors, including formation of H$_2$O (Fig.~\ref{fig:fib}d), SiH$_4$ (Extended Data Fig.~\ref{fig:SI:SiH4}), B2~$\mathrm{Fe}_{1-y}\mathrm{Si}_y$ (Extended Data Fig.~\ref{fig:SI:SiO2}), and FeH$_x$ (Fig.~\ref{fig:xrd}b) (Supplementary Discussion~\ref{SI:section:SiO2-fay-summary}). 
SiH$_4$ was detected in runs in which fayalite was melted in an Ar medium with 50\% H$_2$ (Supplementary Discussion~\ref{SI:section:fay-50H}). In this case, the relative content of H$_2$O (formed from the reaction) compared to H$_2$ is much higher, due to the decreased initial concentration of H and the increased H$_2$O production from the Fe$^{2+}$ in the silicate, resulting in far more oxidizing conditions. 
In SiH$_4$, Si remains oxidized and therefore water production can continue as the water concentration rises through the formation of SiH$_4$ rather than Fe-Si metal (reaction~\ref{eq:ol-Si-dissolve}). 

\section*{Implications of Reaction for Planets}

In a sub-Neptune planet with a rocky core and significant amount of hydrogen, the core-envelope boundary (CEB) is the most likely region to experience reactions between dense fluid hydrogen and silicate melt. 
The hydrogen-magma boundaries of such planets in the mass range of 3--10\,$M_{E}$ with 2--20~wt\% of H + He atmosphere are expected to reach $P$-$T$ conditions similar to our experimental conditions for water production (Supplementary Discussion~\ref{SI:section:PT-sub-Neptune}).
According to our calculation following ref.~\cite{vazan2018Contribution}, for a $5\,M_{E}$ rocky planet with a 5~wt\% H$_2$ envelope, the core's temperature remains sufficiently high to maintain the molten state of silicates for billions of years (Extended Data Fig.~\ref{fig:SI:amb-PT}), which could in turn make the continuation of the observed reactions possible for billions of years. 
Additionally, because pressure dramatically enhances H$_2$ solubility in magma \cite{hirschmann2012solubility} and can result in miscibility between hydrogen fluid and oxide melts \cite{kim2023Stability}, hydrogen can reach greater depths below the CEB\@.  
Vigorous convection in the molten core could also transport hydrogen further to greater depths.
For all of these reasons, the reactions discussed can continue into the deep interior.

\begin{figure}[!ht]
\centering
\includegraphics[width=\textwidth]{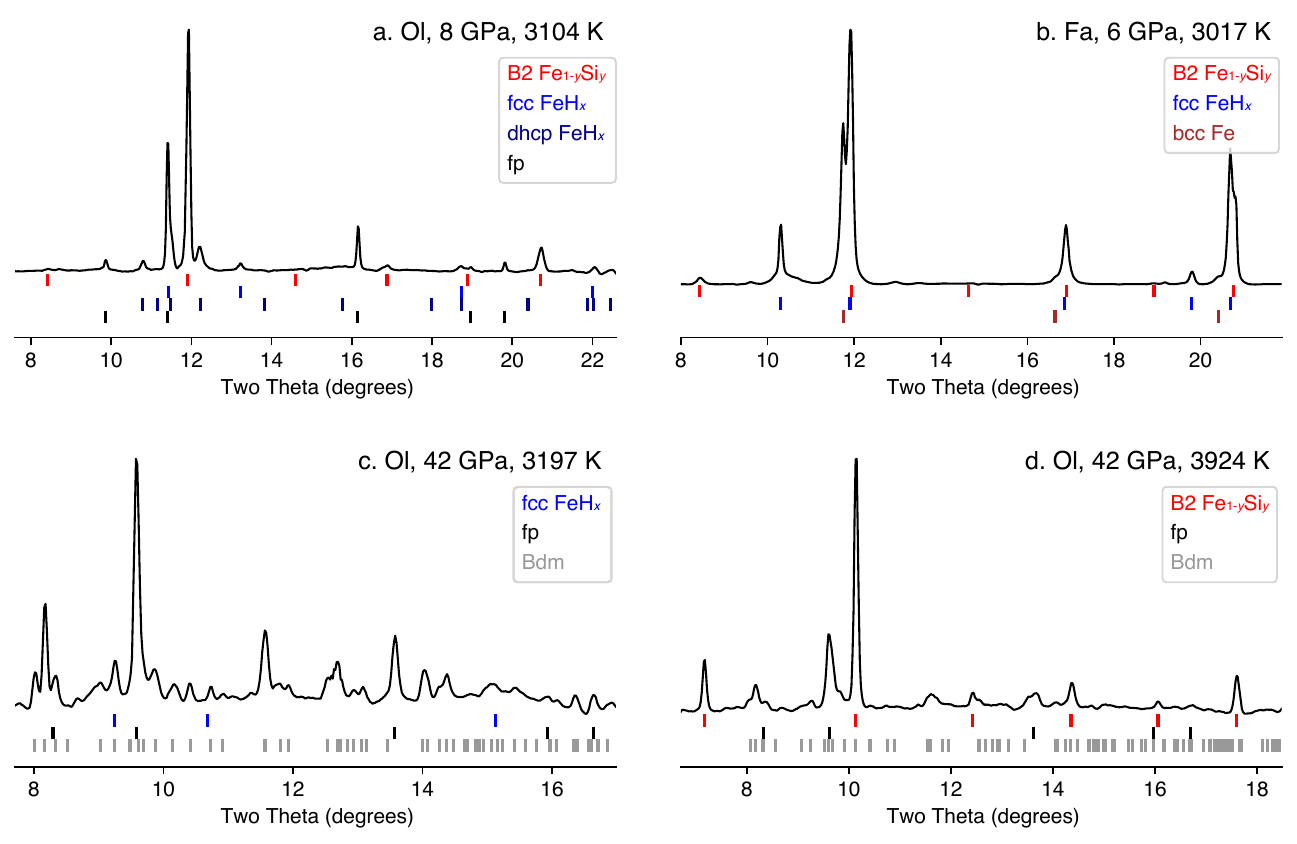}
\caption{{\bf X-ray diffraction patterns after heating silicate samples in a hydrogen medium.} 
{\bf a,} In run SCO-3, olivine breaks down to Fe metal alloys and MgO\@.
{\bf b,} In run FAY-1, fayalite breaks down with Fe and Si present as metal alloys.
A very small amount of bcc~Fe may exist.
{\bf c,} Heating to temperatures below melting (2725--3197~K), bridgmanite (bdm) and ferropericlase (fp) appear at 42~GPa (SCO-13a).
{\bf d,} When the bdm + fp were melted (3352--3924~K at 42~GPa), bdm mostly breaks down and B2~$\mathrm{Fe}_{1-y}\mathrm{Si}_y$ appears (run SCO-13b).
X-ray energy is 30~keV for {\bf b} and 37~keV for {\bf a}, {\bf c}, and {\bf d}. 
The ticks below the diffraction patterns are the peak positions of the observed phases.}
\label{fig:xrd}
\end{figure}

We combined reactions~\ref{eq:ol-Si-reduction} and \ref{eq:ol-Si-dissolve} with constraints from experimental setups and observations to estimate the quantities of reactants and reaction products for olivine reacting with hydrogen in our experiments (Supplementary Discussion~\ref{SI:section:reactant-product-quantities}). 
Considering uncertainties (Supplementary Code~1), we calculated that prior to reacting, the sample volume that would be heated consisted of 4.5--5.7~wt\% H$_2$, $\sim$76~wt\% silicate, and $\sim$19~wt\% Fe metal. 
It shows that 18.1(5)~wt\% H$_2$O was produced. 
This composition lies within the range considered for sub-Neptunes \cite{bean2021nature}.
The complete disappearance of silicate (and therefore release of all of the Si) at lower pressures suggests silicate-undersaturated conditions for the hydrogen-to-silicate mass ratio of our olivine + Fe metal experiment, which was estimated to 0.06--0.08. 
For a sub-Neptune with 4~wt\% H$_2$ and an Earth-like core composition, the hydrogen-to-silicate mass ratio is 0.06.  
Therefore, sub-Neptunes with a few wt\% H$_2$ could provide a silicate undersaturated condition with respect to hydrogen.
At $>$25~GPa, a trace amount of bdm (silicate) remained in the reacted area (Fig.~\ref{fig:xrd}d), which could result from the reverse reaction and therefore suggest a possible equilibrium.  
If this is the case, even under equilibrium conditions, the amount of Si in the silicate melt consumed by the reaction is very large and the H$_2$O produced can amount to tens of weight percent (Supplementary Code~1).  

A theoretical study considering the formation of SiH$_4$ at hydrogen-magma boundaries \cite{misener2023Atmospheresa} predicted $X(\mathrm{H_2O}) \approx 0.2$ (where $X$ is a mole fraction in the envelope) of water production along with $X(\mathrm{SiH}_4) \approx 0.04$, and $X(\mathrm{SiO}) \approx 0.08$ in a $4\,M_{E}$ sub-Neptune with 2.5~wt\% H$_2$ at $\sim$10~GPa and 5000~K\@. %
Another theoretical study\cite{schlichting2022Chemical} considering the reduction of Si predicted $X(\mathrm{H_2O}) \approx 10^{-4}\mathrm{-}10^{-3}$ of water production for a $4\,M_{E}$ sub-Neptune with a CEB temperature of 4500~K\@.
At similar $P$-$T$ conditions, however, our experimental observations indicate much more efficient endogenic water production of $X(\mathrm{H_2O}) = 0.38{-}0.56$, 2--3000 times higher.
While the modeling studies used extrapolation of assumed ideal gas behavior of H$_2$ at much lower pressure, our experiments were conducted at the $P$-$T$ conditions for dense fluid hydrogen, directly relevant to the conditions and states expected at the CEB's of sub-Neptunes.
Comparison of our results with refs~\cite{schlichting2022Chemical,misener2023Atmospheresa} highlights the significant pressure effects on water production.

At a planetary scale, the extent of endogenic water production can be affected by the activity of H$_2$ and H$_2$O in the system.
Under a hydrogen-dominated envelope lacking H$_2$O, strongly reducing conditions promote the reduction of Si as a key pathway for endogenic water production.
As more water is produced (and therefore the activity of H$_2$ decreases), the reaction zone becomes less reducing.
As the reaction progresses, if the water concentration reaches a point where the conditions become sufficiently oxidizing, the reduction of Si could stop.
In this case, SiH$_4$ would become a dominant Si-bearing product of the hydrogen-magma reaction because it does not require the reduction of Si$^{4+}$. 
Because of this, SiH$_4$ formation can occur at more oxidizing conditions than Fe-Si alloys, as shown in our experiments (Supplementary Discussion~\ref{SI:section:fay-50H}), prolonging endogenic water production to more oxidizing conditions. 

\begin{figure}[!ht]
    \centering
    \includegraphics[width=\textwidth]{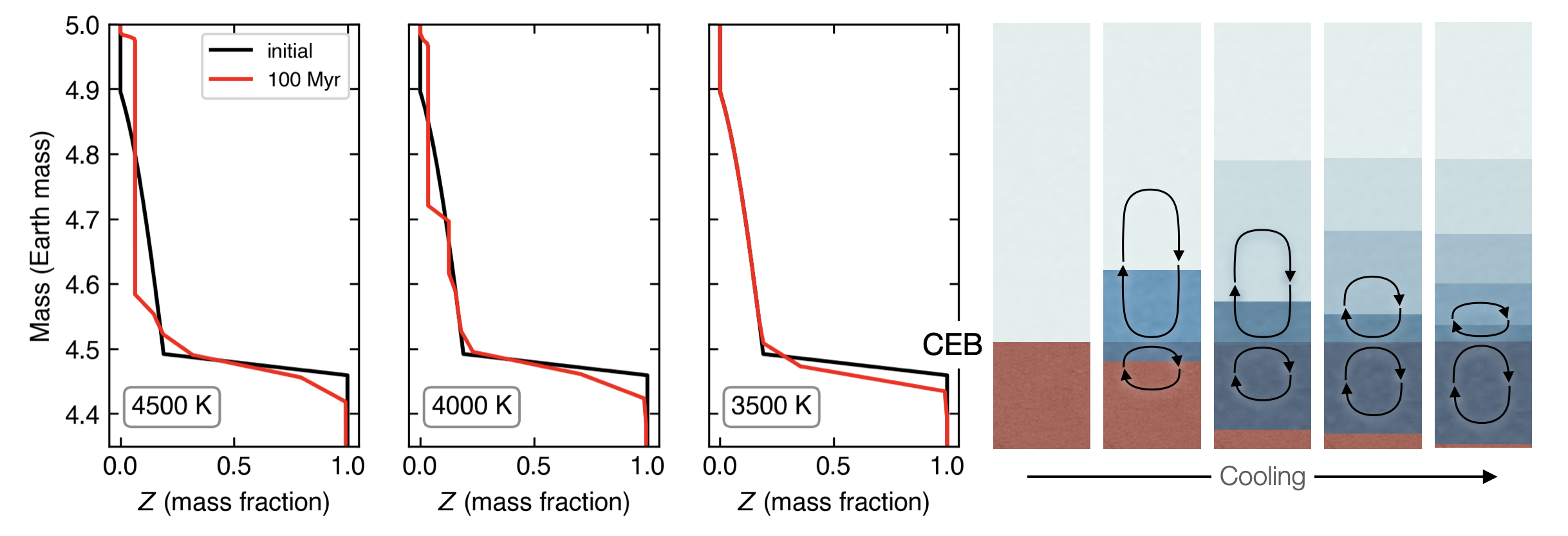} %
    \caption{\textbf{Water redistribution in a $5\,M_{E}$ planet with 10~wt\% envelope made of H$_2$ and He.}
    Shown is the water mass fraction $(Z)$ for an initial state distribution (black) and after 100\,Myr of evolution (red). 
    Mixing efficiency decreases as the planet cools (left to right in the left panel).
    The schematic diagram on the right side shows time progression of mixing scale near the reaction zone.
    }\label{fig:AV2}
\end{figure}

The redox state of the reaction zone can also be affected by the dynamics of the region.
In the absence of mixing and transport, the fraction of water near the reaction zone will increase which will slow down and ultimately stop the water-producing reaction when the water concentration exceeds a critical value (which is not well constrained but we estimate to be $X(\mathrm{H_2O}) \geq 38{-}56$~mol\%; Supplementary Discussion~\ref{SI:section:reactant-product-quantities} and Supplementary Code~1).
However, interiors of sub-Neptunes by and large are convective, and therefore efficient mixing and transport can reduce the concentration of H$_2$O in the reaction zone. 
In Fig.~\ref{fig:AV2}, we show an example of the efficiency of water-mixing over 100 million years in a $5\,M_{E}$ rocky planet with a 10\% gas (H,He) envelope. 
For CEB temperatures of $\geq$4500~K, water is almost completely mixed through convection in the envelope. 
The mixing efficiency decreases as the planet cools. 
Because water production continues up to $X(\mathrm{H_2O}) > 0.3$, we anticipate that convective mixing may stop before water production has ended. 
Consequently, the deep envelope is expected to be more water-rich than the outer envelope in such planets at temperatures lower than $\sim$3500~K\@.

Hydrides also form in the reactions discussed here and in other experiments\cite{kim2023Stability,shinozaki2013Influence,shinozaki2014formation}.
Only a small amount of MgH$_2$ was reported to be produced and its formation is limited to temperatures well over 3000~K (Supplementary Discussion~\ref{SI:section:Mg-reactions}).
Estimates from our experiments (Supplementary Discussion~\ref{SI:section:reactant-product-quantities}) show $6{-}23(5)$~mol\% SiH$_4$ can be produced and therefore it is important to consider.
Although no data exists directly at the $P$-$T$ conditions of our study, existing data indicate that the density of SiH$_4$ is 20--30\% lower than that of H$_2$O at 300~K and relevant pressures (Supplementary Discussion~\ref{SI:section:SiH4-dynamics}).
Therefore, the upward transport of H$_2$O and SiH$_4$ away from the reaction zone could be more efficient than predicted by our model. 

It is important to mention that at the $P$-$T$ conditions we consider here, H$_2$ and H$_2$O are completely miscible and become a single fluid \cite{gupta2025Miscibility} while our dynamic model includes only convective mixing of two separate fluids of H$_2$ and H$_2$O.
Similar to H$_2$O and H$_2$, existing data also supports the miscibility between SiH$_4$ and H$_2$ at the conditions of the CEB\@.
The miscibility will decrease the density of the fluid compared to pure H$_2$O, and the hydrogen contained in the dense fluid layer will still react with the magma ocean (Supplementary Discussion~\ref{SI:section:SiH4-dynamics}).

The increasing solubility of water in silicates at higher pressures (i.e., ingassing) \cite{kim2021atomic,hirschmann2005Storage} can reduce its activity at the CEB and thus prolong the hydrogen-silicate reactions. 
In addition, the dissolved water decreases both the melting temperature and viscosity of the magma, prolonging the molten state of the interior and promoting efficient convective mixing of materials to the deeper under-saturated depths of the core.
Therefore, strong ingassing of water (and hydrogen) to the core can also enable more water production.

The amount of endogenic water produced can vary depending on the properties and configurations of different planets (Supplementary Discussion~\ref{SI:section:Fe-segregation}). 
The Mg/Si ratio can directly affect the amount of water produced, as Si-involved reactions with hydrogen likely dominate endogenic water production compared to Mg and Fe (Supplementary Discussion~\ref{SI:section:Mg-reactions}). 
For example, for a planet with an Earth-like metal-to-silicate mass ratio but the Mg/Si ratio reduced from 2 to 0.5 (more Si), if all the silicate reacts with hydrogen ($>$3~wt\% of H$_2$), the amount of water produced will increase from 16 to 29~wt\% (Supplementary Code~3).
The Mg/Si ratio will also affect the viscosity of the magma and therefore impact the ingassing and mixing of volatile species \cite{karki2018Simulation}. 
Therefore, the large variation in Mg/Si ratios observed in exoplanetary systems \cite{putirka2021Polluted} will result in variation in endogenic water production among planets with rocky interiors and hydrogen-dominated atmospheres. 

The proposed atmospheric loss of a sub-Neptune's hydrogen-rich envelope \cite{owen2013kepler,ginzburg2018Corepowered} may occur concurrently with the endogenic production of water. 
The timing and duration of these two processes will be a critical factor in the amount of water produced.
Atmospheric loss also plays a role in the rate of heat loss, which in turn will influence the duration for which the water-producing reactions can persist because they predominantly occur in molten silicates and metal alloys.

If a sub-Neptune undergoes massive gas loss while water is retained due to its high mean molecular mass \cite{aguichine2021Mass,krissansen-totton2024Erosion}, it is feasible that the limited water production that occurred concurrently with gas loss would result in a rocky planet with surface water.
For example, for a $5~M_{E}$ sub-Neptune, if only the outer 5\% of the rocky core reacts with hydrogen, 2--4~wt\% H$_2$O would be produced (Supplementary Code~2). 
At the same time, large amounts of water can also be stored in the core of sub-Neptunes because pressure increases the solubility of water in magma\cite{kim2021atomic,hirschmann2005Storage}.  
Therefore, even if surface/atmospheric water is lost during gas loss, the water stored in the core can contribute significantly to the formation of a secondary atmosphere and hydrosphere when the interior cools and solidifies where the solubility of water is much lower \cite{vazan2022New,luo2024Interior,krissansen-totton2024Erosion}.
Additionally, because higher pressures significantly enhance water-producing reactions, super-Earths converted from hydrogen-rich sub-Neptunes may contain much more water than smaller rocky planets like Earth.

Another key implication is that hydrogen-rich and water-rich sub-Neptunes do not necessarily form through different processes\cite{venturini2020Nature,burn2024Radius}. 
Instead, the reaction we report here suggests that these planet types may be fundamentally related: hydrogen-rich sub-Neptunes could be the precursors of water-rich sub-Neptunes and super-Earths. 
If an excess, unreacted H$_2$ atmosphere can be retained, sub-Neptunes with an H$_2$-rich atmosphere covering an H$_2$O-rich layer above the core (i.e., hycean worlds) may be quite common \cite{madhusudhan2021Habitability}.  

In conventional planet-formation theory, water-rich planets are believed to form outside of the snow line.
The observation of water-rich sub-Neptunes \cite{luque2022Density,piaulet2022Evidence,piaulet-ghorayeb2024JWST,cherubim2023TOI1695,osborne2023TOI544} in close orbits raises important questions about how they can form. 
Models have been proposed to explain close-in transiting water-rich sub-Neptunes, such as migration of water worlds after their formation outside of the snowline \cite{bitsch2021Dry,izidoro2022Exoplanet,burn2024Radius}. 
Endogenic water production through hydrogen-magma reactions observed in our experiments provides a straightforward process to build water-rich sub-Neptunes inside the snow line: close-in orbiting sub-Neptunes can be H$_2$O-rich planets, as the hydrogen-silicate reaction can convert the H-dominated atmosphere to H$_2$O from the inside out, almost independent of the radial distance of them from the stars.
Water production is feasible and expected in the large population of $3\text{--}10\,M_{E}$ planets that were formed with significant ($>$2\%) gas envelopes. 
On planets with a hotter CEB, vigorous convection and sustained water production are expected to persist longer over their thermal evolution.
Moreover, if mixing of water into the core is efficient, the production of water can persist even when convection in the water-polluted envelope becomes less efficient. 
This result has fundamental implications for planet formation and evolution theories. 
We suggest based on experimental results that planets formed from dry materials can become water-rich (tens of wt\% water) planets without direct accretion of water ice. 
Consequently, detection of a large amount of water in exoplanet atmospheres might not be the optimal evidence for planet migration in the protoplanetary disk.
Our new experimental findings challenge the assumed link between planet formation location and composition.

\bibliography{dis.bib,add.bib}

\begin{thebibliography}{10}
\urlstyle{rm}
\expandafter\ifx\csname url\endcsname\relax
  \def\url#1{\texttt{#1}}\fi
\expandafter\ifx\csname urlprefix\endcsname\relax\def\urlprefix{URL }\fi
\expandafter\ifx\csname doiprefix\endcsname\relax\def\doiprefix{DOI: }\fi
\providecommand{\bibinfo}[2]{#2}
\providecommand{\eprint}[2][]{\url{#2}}

\bibitem{fulton2017california}
\bibinfo{author}{Fulton, B.~J.} \emph{et~al.}
\newblock \bibinfo{journal}{\bibinfo{title}{The {California-Kepler} survey. {III.} a gap in the radius distribution of small planets}}.
\newblock {\emph{\JournalTitle{The Astronomical Journal}}} \textbf{\bibinfo{volume}{154}}, \bibinfo{pages}{109} (\bibinfo{year}{2017}).

\bibitem{bean2021nature}
\bibinfo{author}{Bean, J.~L.}, \bibinfo{author}{Raymond, S.~N.} \& \bibinfo{author}{Owen, J.~E.}
\newblock \bibinfo{journal}{\bibinfo{title}{The nature and origins of sub-neptune size planets}}.
\newblock {\emph{\JournalTitle{Journal of Geophysical Research: Planets}}} \textbf{\bibinfo{volume}{126}}, \bibinfo{pages}{e2020JE006639} (\bibinfo{year}{2021}).

\bibitem{bitsch2021Dry}
\bibinfo{author}{Bitsch, B.} \emph{et~al.}
\newblock \bibinfo{journal}{\bibinfo{title}{Dry or water world? {{How}} the water contents of inner sub-{{Neptunes}} constrain giant planet formation and the location of the water ice line}}.
\newblock {\emph{\JournalTitle{Astronomy \& Astrophysics}}} \textbf{\bibinfo{volume}{649}}, \bibinfo{pages}{L5}, \doiprefix\url{10.1051/0004-6361/202140793} (\bibinfo{year}{2021}).

\bibitem{misener2023Atmospheresa}
\bibinfo{author}{Misener, W.}, \bibinfo{author}{Schlichting, H.~E.} \& \bibinfo{author}{Young, E.~D.}
\newblock \bibinfo{journal}{\bibinfo{title}{Atmospheres as windows into sub-{{Neptune}} interiors: Coupled chemistry and structure of hydrogen--silane--water envelopes}}.
\newblock {\emph{\JournalTitle{Monthly Notices of the Royal Astronomical Society}}} \textbf{\bibinfo{volume}{524}}, \bibinfo{pages}{981--992}, \doiprefix\url{10.1093/mnras/stad1910} (\bibinfo{year}{2023}).

\bibitem{schlichting2022Chemical}
\bibinfo{author}{Schlichting, H.~E.} \& \bibinfo{author}{Young, E.~D.}
\newblock \bibinfo{journal}{\bibinfo{title}{Chemical equilibrium between cores, mantles, and atmospheres of super-{{Earths}} and sub-{{Neptunes}} and implications for their compositions, interiors, and evolution}}.
\newblock {\emph{\JournalTitle{The Planetary Science Journal}}} \textbf{\bibinfo{volume}{3}}, \bibinfo{pages}{127}, \doiprefix\url{10.3847/PSJ/ac68e6} (\bibinfo{year}{2022}).

\bibitem{morbidelli2000Source}
\bibinfo{author}{Morbidelli, A.} \emph{et~al.}
\newblock \bibinfo{journal}{\bibinfo{title}{Source regions and timescales for the delivery of water to the {{Earth}}}}.
\newblock {\emph{\JournalTitle{Meteoritics \& Planetary Science}}} \textbf{\bibinfo{volume}{35}}, \bibinfo{pages}{1309--1320}, \doiprefix\url{10.1111/j.1945-5100.2000.tb01518.x} (\bibinfo{year}{2000}).

\bibitem{ikoma2006constraints}
\bibinfo{author}{Ikoma, M.} \& \bibinfo{author}{Genda, H.}
\newblock \bibinfo{journal}{\bibinfo{title}{Constraints on the mass of a habitable planet with water of nebular origin}}.
\newblock {\emph{\JournalTitle{The Astrophysical Journal}}} \textbf{\bibinfo{volume}{648}}, \bibinfo{pages}{696} (\bibinfo{year}{2006}).

\bibitem{hallis2015evidence}
\bibinfo{author}{Hallis, L.~J.} \emph{et~al.}
\newblock \bibinfo{journal}{\bibinfo{title}{Evidence for primordial water in {Earth's} deep mantle}}.
\newblock {\emph{\JournalTitle{Science}}} \textbf{\bibinfo{volume}{350}}, \bibinfo{pages}{795--797} (\bibinfo{year}{2015}).

\bibitem{young2023Earth}
\bibinfo{author}{Young, E.~D.}, \bibinfo{author}{Shahar, A.} \& \bibinfo{author}{Schlichting, H.~E.}
\newblock \bibinfo{journal}{\bibinfo{title}{Earth shaped by primordial {{H}}{\textsubscript{2}} atmospheres}}.
\newblock {\emph{\JournalTitle{Nature}}} \textbf{\bibinfo{volume}{616}}, \bibinfo{pages}{306--311}, \doiprefix\url{10.1038/s41586-023-05823-0} (\bibinfo{year}{2023}).

\bibitem{howard2012Planet}
\bibinfo{author}{Howard, A.~W.} \emph{et~al.}
\newblock \bibinfo{journal}{\bibinfo{title}{Planet occurrence within 0.25 {{AU}} of solar-type stars from {{{\emph{Kepler}}}}}}.
\newblock {\emph{\JournalTitle{The Astrophysical Journal Supplement Series}}} \textbf{\bibinfo{volume}{201}}, \bibinfo{pages}{15}, \doiprefix\url{10.1088/0067-0049/201/2/15} (\bibinfo{year}{2012}).

\bibitem{owen2013kepler}
\bibinfo{author}{Owen, J.~E.} \& \bibinfo{author}{Wu, Y.}
\newblock \bibinfo{journal}{\bibinfo{title}{Kepler planets: a tale of evaporation}}.
\newblock {\emph{\JournalTitle{The Astrophysical Journal}}} \textbf{\bibinfo{volume}{775}}, \bibinfo{pages}{105} (\bibinfo{year}{2013}).

\bibitem{ginzburg2018Corepowered}
\bibinfo{author}{Ginzburg, S.}, \bibinfo{author}{Schlichting, H.~E.} \& \bibinfo{author}{Sari, R.}
\newblock \bibinfo{journal}{\bibinfo{title}{Core-powered mass-loss and the radius distribution of small exoplanets}}.
\newblock {\emph{\JournalTitle{Monthly Notices of the Royal Astronomical Society}}} \textbf{\bibinfo{volume}{476}}, \bibinfo{pages}{759--765}, \doiprefix\url{10.1093/mnras/sty290} (\bibinfo{year}{2018}).

\bibitem{zeng2019growth}
\bibinfo{author}{Zeng, L.} \emph{et~al.}
\newblock \bibinfo{journal}{\bibinfo{title}{Growth model interpretation of planet size distribution}}.
\newblock {\emph{\JournalTitle{Proceedings of the National Academy of Sciences}}} \textbf{\bibinfo{volume}{116}}, \bibinfo{pages}{9723--9728} (\bibinfo{year}{2019}).

\bibitem{venturini2020jupiter}
\bibinfo{author}{Venturini, J.} \& \bibinfo{author}{Helled, R.}
\newblock \bibinfo{journal}{\bibinfo{title}{Jupiter’s heavy-element enrichment expected from formation models}}.
\newblock {\emph{\JournalTitle{Astronomy \& Astrophysics}}} \textbf{\bibinfo{volume}{634}}, \bibinfo{pages}{A31} (\bibinfo{year}{2020}).

\bibitem{luque2022Density}
\bibinfo{author}{Luque, R.} \& \bibinfo{author}{Pall{\'e}, E.}
\newblock \bibinfo{journal}{\bibinfo{title}{Density, not radius, separates rocky and water-rich small planets orbiting {{M}} dwarf stars}}.
\newblock {\emph{\JournalTitle{Science}}} \textbf{\bibinfo{volume}{377}}, \bibinfo{pages}{1211--1214}, \doiprefix\url{10.1126/science.abl7164} (\bibinfo{year}{2022}).

\bibitem{piaulet2022Evidence}
\bibinfo{author}{Piaulet, C.} \emph{et~al.}
\newblock \bibinfo{journal}{\bibinfo{title}{Evidence for the volatile-rich composition of a 1.5-{{Earth-radius}} planet}}.
\newblock {\emph{\JournalTitle{Nature Astronomy}}} \doiprefix\url{10.1038/s41550-022-01835-4} (\bibinfo{year}{2022}).

\bibitem{piaulet-ghorayeb2024JWST}
\bibinfo{author}{Piaulet-Ghorayeb, C.} \emph{et~al.}
\newblock \bibinfo{journal}{\bibinfo{title}{{JWST}/{NIRISS} {Reveals} the {Water}-rich “{Steam} {World}” {Atmosphere} of {GJ} 9827 d}}.
\newblock {\emph{\JournalTitle{The Astrophysical Journal Letters}}} \textbf{\bibinfo{volume}{974}}, \bibinfo{pages}{L10}, \doiprefix\url{10.3847/2041-8213/ad6f00} (\bibinfo{year}{2024}).

\bibitem{hirschmann2012solubility}
\bibinfo{author}{Hirschmann, M.~M.}, \bibinfo{author}{Withers, A.}, \bibinfo{author}{Ardia, P.} \& \bibinfo{author}{Foley, N.}
\newblock \bibinfo{journal}{\bibinfo{title}{Solubility of molecular hydrogen in silicate melts and consequences for volatile evolution of terrestrial planets}}.
\newblock {\emph{\JournalTitle{Earth and Planetary Science Letters}}} \textbf{\bibinfo{volume}{345}}, \bibinfo{pages}{38--48} (\bibinfo{year}{2012}).

\bibitem{kite2019superabundance}
\bibinfo{author}{Kite, E.~S.}, \bibinfo{author}{Fegley~Jr, B.}, \bibinfo{author}{Schaefer, L.} \& \bibinfo{author}{Ford, E.~B.}
\newblock \bibinfo{journal}{\bibinfo{title}{Superabundance of exoplanet sub-neptunes explained by fugacity crisis}}.
\newblock {\emph{\JournalTitle{The Astrophysical Journal Letters}}} \textbf{\bibinfo{volume}{887}}, \bibinfo{pages}{L33} (\bibinfo{year}{2019}).

\bibitem{sabat2014ReductionOxideMinerals}
\bibinfo{author}{Sabat, K.~C.}, \bibinfo{author}{Rajput, P.}, \bibinfo{author}{Paramguru, R.~K.}, \bibinfo{author}{Bhoi, B.} \& \bibinfo{author}{Mishra, B.~K.}
\newblock \bibinfo{journal}{\bibinfo{title}{Reduction of {{Oxide Minerals}} by {{Hydrogen Plasma}}: {{An Overview}}}}.
\newblock {\emph{\JournalTitle{Plasma Chemistry and Plasma Processing}}} \textbf{\bibinfo{volume}{34}}, \bibinfo{pages}{1--23}, \doiprefix\url{10.1007/s11090-013-9484-2} (\bibinfo{year}{2014}).

\bibitem{kimura2022Predicted}
\bibinfo{author}{Kimura, T.} \& \bibinfo{author}{Ikoma, M.}
\newblock \bibinfo{journal}{\bibinfo{title}{Predicted diversity in water content of terrestrial exoplanets orbiting {{M}} dwarfs}}.
\newblock {\emph{\JournalTitle{Nature Astronomy}}} \textbf{\bibinfo{volume}{6}}, \bibinfo{pages}{1296--1307}, \doiprefix\url{10.1038/s41550-022-01781-1} (\bibinfo{year}{2022}).

\bibitem{krissansen-totton2024Erosion}
\bibinfo{author}{Krissansen-Totton, J.}, \bibinfo{author}{Wogan, N.}, \bibinfo{author}{Thompson, M.} \& \bibinfo{author}{Fortney, J.~J.}
\newblock \bibinfo{journal}{\bibinfo{title}{The erosion of large primary atmospheres typically leaves behind substantial secondary atmospheres on temperate rocky planets}}.
\newblock {\emph{\JournalTitle{Nature Communications}}} \textbf{\bibinfo{volume}{15}}, \bibinfo{pages}{8374}, \doiprefix\url{10.1038/s41467-024-52642-6} (\bibinfo{year}{2024}).

\bibitem{horn2023reaction}
\bibinfo{author}{Horn, H.}, \bibinfo{author}{Prakapenka, V.}, \bibinfo{author}{Chariton, S.}, \bibinfo{author}{Speziale, S.} \& \bibinfo{author}{Shim, S.-H.}
\newblock \bibinfo{journal}{\bibinfo{title}{Reaction between hydrogen and ferrous/ferric oxides at high pressures and high temperatures—implications for sub-neptunes and super-earths}}.
\newblock {\emph{\JournalTitle{The Planetary Science Journal}}} \textbf{\bibinfo{volume}{4}}, \bibinfo{pages}{30} (\bibinfo{year}{2023}).

\bibitem{kim2023Stability}
\bibinfo{author}{Kim, T.} \emph{et~al.}
\newblock \bibinfo{journal}{\bibinfo{title}{Stability of hydrides in sub-{Neptune} exoplanets with thick hydrogen-rich atmospheres}}.
\newblock {\emph{\JournalTitle{Proceedings of the National Academy of Sciences}}} \textbf{\bibinfo{volume}{120}}, \bibinfo{pages}{e2309786120}, \doiprefix\url{10.1073/pnas.2309786120} (\bibinfo{year}{2023}).

\bibitem{shinozaki2013Influence}
\bibinfo{author}{Shinozaki, A.} \emph{et~al.}
\newblock \bibinfo{journal}{\bibinfo{title}{Influence of {H}$_{\textrm{2}}$ fluid on the stability and dissolution of {Mg}$_{\textrm{2}}${SiO}$_{\textrm{4}}$ forsterite under high pressure and high temperature}}.
\newblock {\emph{\JournalTitle{American Mineralogist}}} \textbf{\bibinfo{volume}{98}}, \bibinfo{pages}{1604--1609}, \doiprefix\url{10.2138/am.2013.4434} (\bibinfo{year}{2013}).

\bibitem{shinozaki2014formation}
\bibinfo{author}{Shinozaki, A.} \emph{et~al.}
\newblock \bibinfo{journal}{\bibinfo{title}{Formation of {SiH}$_4$ and {H$_2$O} by the dissolution of quartz in {H$_2$} fluid under high pressure and temperature}}.
\newblock {\emph{\JournalTitle{American Mineralogist}}} \textbf{\bibinfo{volume}{99}}, \bibinfo{pages}{1265--1269} (\bibinfo{year}{2014}).

\bibitem{stokl2016Dynamical}
\bibinfo{author}{St{\"o}kl, A.}, \bibinfo{author}{Dorfi, E.~A.}, \bibinfo{author}{Johnstone, C.~P.} \& \bibinfo{author}{Lammer, H.}
\newblock \bibinfo{journal}{\bibinfo{title}{Dynamical accretion of primordial atmospheres around planets with masses between 0.1 and 5~{$M(E)$} in the habitable zone}}.
\newblock {\emph{\JournalTitle{The Astrophysical Journal}}} \textbf{\bibinfo{volume}{825}}, \bibinfo{pages}{86}, \doiprefix\url{10.3847/0004-637X/825/2/86} (\bibinfo{year}{2016}).

\bibitem{vazan2018Contribution}
\bibinfo{author}{Vazan, A.}, \bibinfo{author}{Ormel, C.~W.}, \bibinfo{author}{Noack, L.} \& \bibinfo{author}{Dominik, C.}
\newblock \bibinfo{journal}{\bibinfo{title}{Contribution of the core to the thermal evolution of sub-{{Neptunes}}}}.
\newblock {\emph{\JournalTitle{The Astrophysical Journal}}} \textbf{\bibinfo{volume}{869}}, \bibinfo{pages}{163}, \doiprefix\url{10.3847/1538-4357/aaef33} (\bibinfo{year}{2018}).

\bibitem{goncharov2010x}
\bibinfo{author}{Goncharov, A.~F.} \emph{et~al.}
\newblock \bibinfo{journal}{\bibinfo{title}{X-ray diffraction in the pulsed laser heated diamond anvil cell}}.
\newblock {\emph{\JournalTitle{Review of Scientific Instruments}}} \textbf{\bibinfo{volume}{81}}, \bibinfo{pages}{113902} (\bibinfo{year}{2010}).

\bibitem{shen1995melting}
\bibinfo{author}{Shen, G.} \& \bibinfo{author}{Lazor, P.}
\newblock \bibinfo{journal}{\bibinfo{title}{Melting of minerals under the lower mantle conditions: experimental results}}.
\newblock {\emph{\JournalTitle{Journal of Geophysical Research: Solid Earth}}} \textbf{\bibinfo{volume}{100}}, \bibinfo{pages}{17699--17713} (\bibinfo{year}{1995}).

\bibitem{gupta2025Miscibility}
\bibinfo{author}{Gupta, A.}, \bibinfo{author}{Stixrude, L.} \& \bibinfo{author}{Schlichting, H.~E.}
\newblock \bibinfo{journal}{\bibinfo{title}{The miscibility of hydrogen and water in planetary atmospheres and interiors}}.
\newblock {\emph{\JournalTitle{The Astrophysical Journal Letters}}} \textbf{\bibinfo{volume}{982}}, \bibinfo{pages}{L35}, \doiprefix\url{10.3847/2041-8213/adb631} (\bibinfo{year}{2025}).

\bibitem{kim2021atomic}
\bibinfo{author}{Kim, T.} \emph{et~al.}
\newblock \bibinfo{journal}{\bibinfo{title}{Atomic-scale mixing between {MgO} and {H$_2$O} in the deep interiors of water-rich planets}}.
\newblock {\emph{\JournalTitle{Nature Astronomy}}} \bibinfo{pages}{1--7} (\bibinfo{year}{2021}).

\bibitem{hirschmann2005Storage}
\bibinfo{author}{Hirschmann, M.~M.}, \bibinfo{author}{Aubaud, C.} \& \bibinfo{author}{Withers, A.~C.}
\newblock \bibinfo{journal}{\bibinfo{title}{Storage capacity of {{H}}{\textsubscript{2}}{{O}} in nominally anhydrous minerals in the upper mantle}}.
\newblock {\emph{\JournalTitle{Earth and Planetary Science Letters}}} \textbf{\bibinfo{volume}{236}}, \bibinfo{pages}{167--181}, \doiprefix\url{10.1016/j.epsl.2005.04.022} (\bibinfo{year}{2005}).

\bibitem{karki2018Simulation}
\bibinfo{author}{Karki, B.~B.}, \bibinfo{author}{Ghosh, D.~B.} \& \bibinfo{author}{Bajgain, S.~K.}
\newblock \bibinfo{title}{Simulation of {{Silicate Melts Under Pressure}}}.
\newblock In \emph{\bibinfo{booktitle}{Magmas {{Under Pressure}}}}, \bibinfo{pages}{419--453}, \doiprefix\url{10.1016/B978-0-12-811301-1.00016-2} (\bibinfo{publisher}{Elsevier}, \bibinfo{year}{2018}).

\bibitem{putirka2021Polluted}
\bibinfo{author}{Putirka, K.~D.} \& \bibinfo{author}{Xu, S.}
\newblock \bibinfo{journal}{\bibinfo{title}{Polluted white dwarfs reveal exotic mantle rock types on exoplanets in our solar neighborhood}}.
\newblock {\emph{\JournalTitle{Nature communications}}} \textbf{\bibinfo{volume}{12}}, \bibinfo{pages}{6168} (\bibinfo{year}{2021}).

\bibitem{aguichine2021Mass}
\bibinfo{author}{Aguichine, A.}, \bibinfo{author}{Mousis, O.}, \bibinfo{author}{Deleuil, M.} \& \bibinfo{author}{Marcq, E.}
\newblock \bibinfo{journal}{\bibinfo{title}{Mass--radius relationships for irradiated ocean planets}}.
\newblock {\emph{\JournalTitle{The Astrophysical Journal}}} \textbf{\bibinfo{volume}{914}}, \bibinfo{pages}{84}, \doiprefix\url{10.3847/1538-4357/abfa99} (\bibinfo{year}{2021}).

\bibitem{vazan2022New}
\bibinfo{author}{Vazan, A.}, \bibinfo{author}{Sari, R.} \& \bibinfo{author}{Kessel, R.}
\newblock \bibinfo{journal}{\bibinfo{title}{A new perspective on interiors of ice-rich planets: {{Ice-rock}} mixture instead of ice on top of rock}}.
\newblock {\emph{\JournalTitle{The Astrophysical Journal}}} \textbf{\bibinfo{volume}{926}}, \bibinfo{pages}{150}, \doiprefix\url{10.3847/1538-4357/ac458c} (\bibinfo{year}{2022}).
\newblock \eprint{2011.00602}.

\bibitem{luo2024Interior}
\bibinfo{author}{Luo, H.}
\newblock \bibinfo{journal}{\bibinfo{title}{The interior as the dominant water reservoir in super-{Earths} and sub-{Neptunes}}}.
\newblock {\emph{\JournalTitle{Nature Astronomy}}} \doiprefix\url{https://doi.org/10.1038/s41550-024-02347-z} (\bibinfo{year}{2024}).

\bibitem{venturini2020Nature}
\bibinfo{author}{Venturini, J.}, \bibinfo{author}{Guilera, O.~M.}, \bibinfo{author}{Haldemann, J.}, \bibinfo{author}{Ronco, M.~P.} \& \bibinfo{author}{Mordasini, C.}
\newblock \bibinfo{journal}{\bibinfo{title}{The nature of the radius valley: {{Hints}} from formation and evolution models}}.
\newblock {\emph{\JournalTitle{Astronomy \& Astrophysics}}} \textbf{\bibinfo{volume}{643}}, \bibinfo{pages}{L1}, \doiprefix\url{10.1051/0004-6361/202039141} (\bibinfo{year}{2020}).

\bibitem{burn2024Radius}
\bibinfo{author}{Burn, R.} \emph{et~al.}
\newblock \bibinfo{journal}{\bibinfo{title}{A radius valley between migrated steam worlds and evaporated rocky cores}}.
\newblock {\emph{\JournalTitle{Nature Astronomy}}} \doiprefix\url{10.1038/s41550-023-02183-7} (\bibinfo{year}{2024}).

\bibitem{madhusudhan2021Habitability}
\bibinfo{author}{Madhusudhan, N.}, \bibinfo{author}{Piette, A. A.~A.} \& \bibinfo{author}{Constantinou, S.}
\newblock \bibinfo{journal}{\bibinfo{title}{Habitability and biosignatures of hycean worlds}}.
\newblock {\emph{\JournalTitle{The Astrophysical Journal}}} \textbf{\bibinfo{volume}{918}}, \bibinfo{pages}{1}, \doiprefix\url{10.3847/1538-4357/abfd9c} (\bibinfo{year}{2021}).

\bibitem{cherubim2023TOI1695}
\bibinfo{author}{Cherubim, C.} \emph{et~al.}
\newblock \bibinfo{journal}{\bibinfo{title}{{{TOI-1695}} b: {{A Water World Orbiting}} an {{Early-M Dwarf}} in the {{Planet Radius Valley}}}}.
\newblock {\emph{\JournalTitle{The Astronomical Journal}}} \textbf{\bibinfo{volume}{165}}, \bibinfo{pages}{167}, \doiprefix\url{10.3847/1538-3881/acbdfd} (\bibinfo{year}{2023}).

\bibitem{osborne2023TOI544}
\bibinfo{author}{Osborne, H. L.~M.} \emph{et~al.}
\newblock \bibinfo{journal}{\bibinfo{title}{{{TOI-544}} b: A potential water-world inside the radius valley in a two-planet system}}.
\newblock {\emph{\JournalTitle{Monthly Notices of the Royal Astronomical Society}}} \textbf{\bibinfo{volume}{527}}, \bibinfo{pages}{11138--11157}, \doiprefix\url{10.1093/mnras/stad3837} (\bibinfo{year}{2023}).

\bibitem{izidoro2022Exoplanet}
\bibinfo{author}{Izidoro, A.} \emph{et~al.}
\newblock \bibinfo{journal}{\bibinfo{title}{The exoplanet radius valley from gas-driven planet migration and breaking of resonant chains}}.
\newblock {\emph{\JournalTitle{The Astrophysical Journal Letters}}} \textbf{\bibinfo{volume}{939}}, \bibinfo{pages}{L19}, \doiprefix\url{10.3847/2041-8213/ac990d} (\bibinfo{year}{2022}).

\end{thebibliography}


\begin{thebibliography}{10}
\urlstyle{rm}
\expandafter\ifx\csname url\endcsname\relax
  \def\url#1{\texttt{#1}}\fi
\expandafter\ifx\csname urlprefix\endcsname\relax\def\urlprefix{URL }\fi
\expandafter\ifx\csname doiprefix\endcsname\relax\def\doiprefix{DOI: }\fi
\providecommand{\bibinfo}[2]{#2}
\providecommand{\eprint}[2][]{\url{#2}}

\bibitem{piermarini1975calibration}
\bibinfo{author}{Piermarini, G.~J.}, \bibinfo{author}{Block, S.}, \bibinfo{author}{Barnett, J.} \& \bibinfo{author}{Forman, R.}
\newblock \bibinfo{journal}{\bibinfo{title}{Calibration of the pressure dependence of the r 1 ruby fluorescence line to 195 kbar}}.
\newblock {\emph{\JournalTitle{Journal of Applied Physics}}} \textbf{\bibinfo{volume}{46}}, \bibinfo{pages}{2774--2780} (\bibinfo{year}{1975}).

\bibitem{prakapenka2008advanced}
\bibinfo{author}{Prakapenka, V.} \emph{et~al.}
\newblock \bibinfo{journal}{\bibinfo{title}{Advanced flat top laser heating system for high pressure research at {GSECARS}: application to the melting behavior of germanium}}.
\newblock {\emph{\JournalTitle{High Pressure Research}}} \textbf{\bibinfo{volume}{28}}, \bibinfo{pages}{225--235} (\bibinfo{year}{2008}).

\bibitem{deemyad2005pulsed}
\bibinfo{author}{Deemyad, S.} \emph{et~al.}
\newblock \bibinfo{journal}{\bibinfo{title}{Pulsed laser heating and temperature determination in a diamond anvil cell}}.
\newblock {\emph{\JournalTitle{Review of Scientific Instruments}}} \textbf{\bibinfo{volume}{76}}, \bibinfo{pages}{125104} (\bibinfo{year}{2005}).

\bibitem{fu2022stable}
\bibinfo{author}{Fu, S.}, \bibinfo{author}{Chariton, S.}, \bibinfo{author}{Prakapenka, V.~B.}, \bibinfo{author}{Chizmeshya, A.} \& \bibinfo{author}{Shim, S.-H.}
\newblock \bibinfo{journal}{\bibinfo{title}{Stable hexagonal ternary alloy phase in {Fe}-{Si}-{H} at 28.6--42.2~{GPa} and 3000~{K}}}.
\newblock {\emph{\JournalTitle{Physical Review B}}} \textbf{\bibinfo{volume}{105}}, \bibinfo{pages}{104111} (\bibinfo{year}{2022}).

\bibitem{fu2023CoreOriginSeismic}
\bibinfo{author}{Fu, S.}, \bibinfo{author}{Chariton, S.}, \bibinfo{author}{Prakapenka, V.~B.} \& \bibinfo{author}{Shim, S.-H.}
\newblock \bibinfo{journal}{\bibinfo{title}{Core origin of seismic velocity anomalies at {{Earth}}'s core\textendash mantle boundary}}.
\newblock {\emph{\JournalTitle{Nature}}} \doiprefix\url{10.1038/s41586-023-05713-5} (\bibinfo{year}{2023}).

\bibitem{kulka2020Bridgmanite}
\bibinfo{author}{Kulka, B.~L.}, \bibinfo{author}{Dolinschi, J.~D.}, \bibinfo{author}{Leinenweber, K.~D.}, \bibinfo{author}{Prakapenka, V.~B.} \& \bibinfo{author}{Shim, S.-H.}
\newblock \bibinfo{journal}{\bibinfo{title}{The bridgmanite\textendash akimotoite\textendash majorite triple point determined in large volume press and laser-heated diamond anvil cell}}.
\newblock {\emph{\JournalTitle{Minerals}}} \textbf{\bibinfo{volume}{10}}, \bibinfo{pages}{67}, \doiprefix\url{10.3390/min10010067} (\bibinfo{year}{2020}).

\bibitem{prescher2015dioptas}
\bibinfo{author}{Prescher, C.} \& \bibinfo{author}{Prakapenka, V.~B.}
\newblock \bibinfo{journal}{\bibinfo{title}{{{DIOPTAS}}: A program for reduction of two-dimensional {{X-ray}} diffraction data and data exploration}}.
\newblock {\emph{\JournalTitle{High Pressure Research}}} \textbf{\bibinfo{volume}{35}}, \bibinfo{pages}{223--230} (\bibinfo{year}{2015}).

\bibitem{PeakPo}
\bibinfo{author}{Shim, S.-H.}
\newblock \bibinfo{title}{Peakpo - a python software for x-ray diffraction analysis at high pressure and high temperature}, \doiprefix\url{10.5281/ZENODO.3376238} (\bibinfo{year}{2019}).

\bibitem{ye2017intercomparison}
\bibinfo{author}{Ye, Y.}, \bibinfo{author}{Prakapenka, V.}, \bibinfo{author}{Meng, Y.} \& \bibinfo{author}{Shim, S.-H.}
\newblock \bibinfo{journal}{\bibinfo{title}{Intercomparison of the gold, platinum, and {MgO} pressure scales up to 140~{GPa} and 2500~{K}}}.
\newblock {\emph{\JournalTitle{Journal of Geophysical Research: Solid Earth}}} \textbf{\bibinfo{volume}{122}}, \bibinfo{pages}{3450--3464} (\bibinfo{year}{2017}).

\bibitem{dewaele1998temperature}
\bibinfo{author}{Dewaele, A.}, \bibinfo{author}{Fiquet, G.} \& \bibinfo{author}{Gillet, P.}
\newblock \bibinfo{journal}{\bibinfo{title}{Temperature and pressure distribution in the laser-heated diamond--anvil cell}}.
\newblock {\emph{\JournalTitle{Review of scientific instruments}}} \textbf{\bibinfo{volume}{69}}, \bibinfo{pages}{2421--2426} (\bibinfo{year}{1998}).

\bibitem{holtgrewe2019advanced}
\bibinfo{author}{Holtgrewe, N.}, \bibinfo{author}{Greenberg, E.}, \bibinfo{author}{Prescher, C.}, \bibinfo{author}{Prakapenka, V.~B.} \& \bibinfo{author}{Goncharov, A.~F.}
\newblock \bibinfo{journal}{\bibinfo{title}{Advanced integrated optical spectroscopy system for diamond anvil cell studies at {GSECARS}}}.
\newblock {\emph{\JournalTitle{High Pressure Research}}} \textbf{\bibinfo{volume}{39}}, \bibinfo{pages}{457--470} (\bibinfo{year}{2019}).

\bibitem{vazan2015convection}
\bibinfo{author}{Vazan, A.}, \bibinfo{author}{Helled, R.}, \bibinfo{author}{Kovetz, A.} \& \bibinfo{author}{Podolak, M.}
\newblock \bibinfo{journal}{\bibinfo{title}{Convection and mixing in giant planet evolution}}.
\newblock {\emph{\JournalTitle{The Astrophysical Journal}}} \textbf{\bibinfo{volume}{803}}, \bibinfo{pages}{32} (\bibinfo{year}{2015}).

\bibitem{saumon1995equation}
\bibinfo{author}{Saumon, D.}, \bibinfo{author}{Chabrier, G.} \& \bibinfo{author}{van Horn, H.~M.}
\newblock \bibinfo{journal}{\bibinfo{title}{An equation of state for low-mass stars and giant planets}}.
\newblock {\emph{\JournalTitle{Astrophysical Journal Supplement v. 99, p. 713}}} \textbf{\bibinfo{volume}{99}}, \bibinfo{pages}{713} (\bibinfo{year}{1995}).

\bibitem{vazan2013effect}
\bibinfo{author}{Vazan, A.}, \bibinfo{author}{Kovetz, A.}, \bibinfo{author}{Podolak, M.} \& \bibinfo{author}{Helled, R.}
\newblock \bibinfo{journal}{\bibinfo{title}{The effect of composition on the evolution of giant and intermediate-mass planets}}.
\newblock {\emph{\JournalTitle{Monthly Notices of the Royal Astronomical Society}}} \textbf{\bibinfo{volume}{434}}, \bibinfo{pages}{3283--3292} (\bibinfo{year}{2013}).

\bibitem{freedman2014gaseous}
\bibinfo{author}{Freedman, R.~S.} \emph{et~al.}
\newblock \bibinfo{journal}{\bibinfo{title}{Gaseous mean opacities for giant planet and ultracool dwarf atmospheres over a range of metallicities and temperatures}}.
\newblock {\emph{\JournalTitle{The Astrophysical Journal Supplement Series}}} \textbf{\bibinfo{volume}{214}}, \bibinfo{pages}{25} (\bibinfo{year}{2014}).

\bibitem{shim2025Jupyter}
\bibinfo{author}{Shim, S.-H.}
\newblock \bibinfo{title}{Jupyter notebooks for {{Supplementary Codes}}}.
\newblock \bibinfo{howpublished}{Zenodo}, \doiprefix\url{10.5281/ZENODO.15678598} (\bibinfo{year}{2025}).

\bibitem{sakamaki2009}
\bibinfo{author}{Sakamaki, K.} \emph{et~al.}
\newblock \bibinfo{journal}{\bibinfo{title}{Melting phase relation of {FeH$_x$} up to 20~{GPa}: Implication for the temperature of the {Earth's} core}}.
\newblock {\emph{\JournalTitle{Physics of the Earth and Planetary Interiors}}} \textbf{\bibinfo{volume}{174}}, \bibinfo{pages}{192--201} (\bibinfo{year}{2009}).

\bibitem{mosenfelder2007thermodynamic}
\bibinfo{author}{Mosenfelder, J.~L.}, \bibinfo{author}{Asimow, P.~D.} \& \bibinfo{author}{Ahrens, T.~J.}
\newblock \bibinfo{journal}{\bibinfo{title}{Thermodynamic properties of {Mg$_2$SiO$_4$} liquid at ultra-high pressures from shock measurements to 200~{GPa} on forsterite and wadsleyite}}.
\newblock {\emph{\JournalTitle{Journal of Geophysical Research: Solid Earth}}} \textbf{\bibinfo{volume}{112}} (\bibinfo{year}{2007}).

\bibitem{ohtani1979melting}
\bibinfo{author}{Ohtani, E.}
\newblock \bibinfo{journal}{\bibinfo{title}{Melting relation of {Fe$_2$SiO$_4$} up to about 200~kbar}}.
\newblock {\emph{\JournalTitle{Journal of Physics of the Earth}}} \textbf{\bibinfo{volume}{27}}, \bibinfo{pages}{189--208} (\bibinfo{year}{1979}).

\bibitem{andrault2020melting}
\bibinfo{author}{Andrault, D.} \emph{et~al.}
\newblock \bibinfo{journal}{\bibinfo{title}{Melting behavior of {SiO$_2$} up to 120~{GPa}}}.
\newblock {\emph{\JournalTitle{Physics and Chemistry of Minerals}}} \textbf{\bibinfo{volume}{47}}, \bibinfo{pages}{1--9} (\bibinfo{year}{2020}).

\bibitem{zha2017Melting}
\bibinfo{author}{Zha, C.-s.}, \bibinfo{author}{Liu, H.}, \bibinfo{author}{Tse, J.~S.} \& \bibinfo{author}{Hemley, R.~J.}
\newblock \bibinfo{journal}{\bibinfo{title}{Melting and high {{P}}-{{T}} transitions of hydrogen up to 300 {{GPa}}}}.
\newblock {\emph{\JournalTitle{Physical Review Letters}}} \textbf{\bibinfo{volume}{119}}, \bibinfo{pages}{075302}, \doiprefix\url{10.1103/PhysRevLett.119.075302} (\bibinfo{year}{2017}).

\bibitem{narygina2011x}
\bibinfo{author}{Narygina, O.} \emph{et~al.}
\newblock \bibinfo{journal}{\bibinfo{title}{X-ray diffraction and {M\"o}ssbauer spectroscopy study of fcc iron hydride {FeH} at high pressures and implications for the composition of the {Earth's} core}}.
\newblock {\emph{\JournalTitle{Earth and Planetary Science Letters}}} \textbf{\bibinfo{volume}{307}}, \bibinfo{pages}{409--414} (\bibinfo{year}{2011}).

\bibitem{thompson2018high}
\bibinfo{author}{Thompson, E.} \emph{et~al.}
\newblock \bibinfo{journal}{\bibinfo{title}{High-pressure geophysical properties of fcc phase {FeH$_X$}}}.
\newblock {\emph{\JournalTitle{Geochemistry, Geophysics, Geosystems}}} \textbf{\bibinfo{volume}{19}}, \bibinfo{pages}{305--314} (\bibinfo{year}{2018}).

\bibitem{kato2020hydride}
\bibinfo{author}{Kato, C.} \emph{et~al.}
\newblock \bibinfo{journal}{\bibinfo{title}{Stability of fcc phase {FeH} to 137~{GPa}}}.
\newblock {\emph{\JournalTitle{American Mineralogist}}} \textbf{\bibinfo{volume}{105}}, \bibinfo{pages}{917--921} (\bibinfo{year}{2020}).

\bibitem{tagawa2022HighTemperature}
\bibinfo{author}{Tagawa, S.}, \bibinfo{author}{Gomi, H.}, \bibinfo{author}{Hirose, K.} \& \bibinfo{author}{Ohishi, Y.}
\newblock \bibinfo{journal}{\bibinfo{title}{High-{{Temperature Equation}} of {{State}} of {{FeH}}: {{Implications}} for {{Hydrogen}} in {{Earth}}'s {{Inner Core}}}}.
\newblock {\emph{\JournalTitle{Geophysical Research Letters}}} \textbf{\bibinfo{volume}{49}}, \doiprefix\url{10.1029/2021GL096260} (\bibinfo{year}{2022}).

\bibitem{ikuta2019interstitial}
\bibinfo{author}{Ikuta, D.} \emph{et~al.}
\newblock \bibinfo{journal}{\bibinfo{title}{Interstitial hydrogen atoms in face-centered cubic iron in the {Earth's} core}}.
\newblock {\emph{\JournalTitle{Scientific reports}}} \textbf{\bibinfo{volume}{9}}, \bibinfo{pages}{1--8} (\bibinfo{year}{2019}).

\bibitem{shibazaki2014high}
\bibinfo{author}{Shibazaki, Y.} \emph{et~al.}
\newblock \bibinfo{journal}{\bibinfo{title}{High-pressure and high-temperature phase diagram for {Fe$_{0.9}$Ni$_{0.1}$--H} alloy}}.
\newblock {\emph{\JournalTitle{Physics of the Earth and Planetary Interiors}}} \textbf{\bibinfo{volume}{228}}, \bibinfo{pages}{192--201} (\bibinfo{year}{2014}).

\bibitem{ohta2019electrical}
\bibinfo{author}{Ohta, K.}, \bibinfo{author}{Suehiro, S.}, \bibinfo{author}{Hirose, K.} \& \bibinfo{author}{Ohishi, Y.}
\newblock \bibinfo{journal}{\bibinfo{title}{Electrical resistivity of fcc phase iron hydrides at high pressures and temperatures}}.
\newblock {\emph{\JournalTitle{Comptes Rendus Geoscience}}} \textbf{\bibinfo{volume}{351}}, \bibinfo{pages}{147--153} (\bibinfo{year}{2019}).

\bibitem{dorogokupets2017Thermodynamics}
\bibinfo{author}{Dorogokupets, P.~I.}, \bibinfo{author}{Dymshits, A.~M.}, \bibinfo{author}{Litasov, K.~D.} \& \bibinfo{author}{Sokolova, T.~S.}
\newblock \bibinfo{journal}{\bibinfo{title}{Thermodynamics and {{Equations}} of {{State}} of {{Iron}} to 350 {{GPa}} and 6000 {{K}}}}.
\newblock {\emph{\JournalTitle{Scientific Reports}}} \textbf{\bibinfo{volume}{7}}, \bibinfo{pages}{41863}, \doiprefix\url{10.1038/srep41863} (\bibinfo{year}{2017}).

\bibitem{piet2023}
\bibinfo{author}{Piet, H.} \emph{et~al.}
\newblock \bibinfo{journal}{\bibinfo{title}{Superstoichiometric {{Alloying}} of {{H}} and {{Close}}-{{Packed Fe}}-{{Ni Metal Under High Pressures}}: {{Implications}} for {{Hydrogen Storage}} in {{Planetary Core}}}}.
\newblock {\emph{\JournalTitle{Geophysical Research Letters}}} \textbf{\bibinfo{volume}{50}}, \doiprefix\url{10.1029/2022GL101155} (\bibinfo{year}{2023}).

\end{thebibliography}


\begin{thebibliography}{10}
\urlstyle{rm}
\expandafter\ifx\csname url\endcsname\relax
  \def\url#1{\texttt{#1}}\fi
\expandafter\ifx\csname urlprefix\endcsname\relax\def\urlprefix{URL }\fi
\expandafter\ifx\csname doiprefix\endcsname\relax\def\doiprefix{DOI: }\fi
\providecommand{\bibinfo}[2]{#2}
\providecommand{\eprint}[2][]{\url{#2}}

\bibitem{ji2019ultrahigh}
\bibinfo{author}{Ji, C.} \emph{et~al.}
\newblock \bibinfo{journal}{\bibinfo{title}{Ultrahigh-pressure isostructural electronic transitions in hydrogen}}.
\newblock {\emph{\JournalTitle{Nature}}} \textbf{\bibinfo{volume}{573}}, \bibinfo{pages}{558--562} (\bibinfo{year}{2019}).

\bibitem{fischer2014equations}
\bibinfo{author}{Fischer, R.~A.} \emph{et~al.}
\newblock \bibinfo{journal}{\bibinfo{title}{Equations of state in the {Fe-FeSi} system at high pressures and temperatures}}.
\newblock {\emph{\JournalTitle{Journal of Geophysical Research: Solid Earth}}} \textbf{\bibinfo{volume}{119}}, \bibinfo{pages}{2810--2827} (\bibinfo{year}{2014}).

\bibitem{fu2022fesiH}
\bibinfo{author}{Fu, S.}, \bibinfo{author}{Chariton, S.}, \bibinfo{author}{Prakapenka, V.~B.}, \bibinfo{author}{Chizmeshya, A.} \& \bibinfo{author}{Shim, S.-H.}
\newblock \bibinfo{journal}{\bibinfo{title}{Hydrogen solubility in {FeSi} alloy phases at high pressures and temperatures}}.
\newblock {\emph{\JournalTitle{American Minerologist}}} \textbf{\bibinfo{volume}{in press}}, \doiprefix\url{https://doi.org/10.2138/am-2022-8295} (\bibinfo{year}{2022}).

\bibitem{slater1964atomic}
\bibinfo{author}{Slater, J.~C.}
\newblock \bibinfo{journal}{\bibinfo{title}{Atomic radii in crystals}}.
\newblock {\emph{\JournalTitle{The Journal of Chemical Physics}}} \textbf{\bibinfo{volume}{41}}, \bibinfo{pages}{3199--3204} (\bibinfo{year}{1964}).

\bibitem{badding1991}
\bibinfo{author}{Badding, J.}, \bibinfo{author}{Hemley, R.} \& \bibinfo{author}{Mao, H.}
\newblock \bibinfo{journal}{\bibinfo{title}{High-pressure chemistry of hydrogen in metals: In situ study of iron hydride}}.
\newblock {\emph{\JournalTitle{Science}}} \textbf{\bibinfo{volume}{253}}, \bibinfo{pages}{421--424} (\bibinfo{year}{1991}).

\bibitem{narygina2011}
\bibinfo{author}{Narygina, O.} \emph{et~al.}
\newblock \bibinfo{journal}{\bibinfo{title}{X-ray diffraction and m{\"o}ssbauer spectroscopy study of fcc iron hydride {FeH} at high pressures and implications for the composition of the {Earth's} core}}.
\newblock {\emph{\JournalTitle{Earth and Planetary Science Letters}}} \textbf{\bibinfo{volume}{307}}, \bibinfo{pages}{409--414} (\bibinfo{year}{2011}).

\bibitem{hirao2004compression}
\bibinfo{author}{Hirao, N.}, \bibinfo{author}{Kondo, T.}, \bibinfo{author}{Ohtani, E.}, \bibinfo{author}{Takemura, K.} \& \bibinfo{author}{Kikegawa, T.}
\newblock \bibinfo{journal}{\bibinfo{title}{Compression of iron hydride to 80~{GPa} and hydrogen in the {Earth's} inner core}}.
\newblock {\emph{\JournalTitle{Geophysical Research Letters}}} \textbf{\bibinfo{volume}{31}} (\bibinfo{year}{2004}).

\bibitem{Jain2013}
\bibinfo{author}{Jain, A.} \emph{et~al.}
\newblock \bibinfo{journal}{\bibinfo{title}{{The Materials Project: A materials genome approach to accelerating materials innovation}}}.
\newblock {\emph{\JournalTitle{APL Materials}}} \textbf{\bibinfo{volume}{1}}, \bibinfo{pages}{011002}, \doiprefix\url{10.1063/1.4812323} (\bibinfo{year}{2013}).

\bibitem{rotter1966ultrasonic}
\bibinfo{author}{Rotter, C.~A.} \& \bibinfo{author}{Smith, C.~S.}
\newblock \bibinfo{journal}{\bibinfo{title}{Ultrasonic equation of state of iron: I. low pressure, room temperature}}.
\newblock {\emph{\JournalTitle{J. Phys. Chem. Solids}}} \textbf{\bibinfo{volume}{27}}, \bibinfo{pages}{197} (\bibinfo{year}{1966}).

\bibitem{utsumi1998volume}
\bibinfo{author}{Utsumi, W.}, \bibinfo{author}{Weidner, D.~J.} \& \bibinfo{author}{Liebermann, R.~C.}
\newblock \bibinfo{journal}{\bibinfo{title}{Volume measurement of {MgO} at high pressures and high temperatures}}.
\newblock {\emph{\JournalTitle{Geophysical Monograph-American Geophysical Union}}} \textbf{\bibinfo{volume}{101}}, \bibinfo{pages}{327--334} (\bibinfo{year}{1998}).

\bibitem{mcguire2017isothermal}
\bibinfo{author}{McGuire, C.}, \bibinfo{author}{Santamaria-P{\'e}rez, D.}, \bibinfo{author}{Makhluf, A.} \& \bibinfo{author}{Kavner, A.}
\newblock \bibinfo{journal}{\bibinfo{title}{Isothermal equation of state and phase stability of {Fe$_5$Si$_3$} up to 96~{GPa} and 3000~{K}}}.
\newblock {\emph{\JournalTitle{Journal of Geophysical Research: Solid Earth}}} \textbf{\bibinfo{volume}{122}}, \bibinfo{pages}{4328--4335} (\bibinfo{year}{2017}).

\bibitem{errandonea2008structural}
\bibinfo{author}{Errandonea, D.} \emph{et~al.}
\newblock \bibinfo{journal}{\bibinfo{title}{Structural stability of {Fe$_5$Si$_3$} and {Ni$_2$Si} studied by high-pressure x-ray diffraction and ab initio total-energy calculations}}.
\newblock {\emph{\JournalTitle{Physical Review B}}} \textbf{\bibinfo{volume}{77}}, \bibinfo{pages}{094113} (\bibinfo{year}{2008}).

\bibitem{fei1994situ}
\bibinfo{author}{Fei, Y.} \& \bibinfo{author}{Mao, H.-k.}
\newblock \bibinfo{journal}{\bibinfo{title}{In situ determination of the nias phase of {FeO} at high pressure and temperature}}.
\newblock {\emph{\JournalTitle{Science}}} \textbf{\bibinfo{volume}{266}}, \bibinfo{pages}{1678--1680} (\bibinfo{year}{1994}).

\bibitem{mao1991effect}
\bibinfo{author}{Mao, H.} \emph{et~al.}
\newblock \bibinfo{journal}{\bibinfo{title}{Effect of pressure, temperature, and composition on lattice parameters and density of {(Fe,Mg)SiO$_3$}-perovskites to 30~{GPa}}}.
\newblock {\emph{\JournalTitle{Journal of Geophysical Research: Solid Earth}}} \textbf{\bibinfo{volume}{96}}, \bibinfo{pages}{8069--8079} (\bibinfo{year}{1991}).

\bibitem{lin2002iron}
\bibinfo{author}{Lin, J.-F.}, \bibinfo{author}{Heinz, D.~L.}, \bibinfo{author}{Campbell, A.~J.}, \bibinfo{author}{Devine, J.~M.} \& \bibinfo{author}{Shen, G.}
\newblock \bibinfo{journal}{\bibinfo{title}{Iron-silicon alloy in {Earth's} core?}}
\newblock {\emph{\JournalTitle{Science}}} \textbf{\bibinfo{volume}{295}}, \bibinfo{pages}{313--315} (\bibinfo{year}{2002}).

\bibitem{akimoto1977high}
\bibinfo{author}{Akimoto, S.}, \bibinfo{author}{Yagi, T.} \& \bibinfo{author}{Inoue, K.}
\newblock \bibinfo{title}{High temperature-pressure phase boundaries in silicate systems using in situ x-ray diffraction}.
\newblock In \emph{\bibinfo{booktitle}{High-Pressure Research}}, \bibinfo{pages}{585--602} (\bibinfo{publisher}{Elsevier}, \bibinfo{year}{1977}).

\bibitem{sinclair1978single}
\bibinfo{author}{Sinclair, W.} \& \bibinfo{author}{Ringwood, A.}
\newblock \bibinfo{journal}{\bibinfo{title}{Single crystal analysis of the structure of stishovite}}.
\newblock {\emph{\JournalTitle{Nature}}} \textbf{\bibinfo{volume}{272}}, \bibinfo{pages}{714--715} (\bibinfo{year}{1978}).

\bibitem{nisr2020}
\bibinfo{author}{Nisr, C.} \emph{et~al.}
\newblock \bibinfo{journal}{\bibinfo{title}{Large {H$_2$O} solubility in dense silica and its implications for the interiors of water-rich planets}}.
\newblock {\emph{\JournalTitle{Proceedings of the National Academy of Sciences}}} \textbf{\bibinfo{volume}{117}}, \bibinfo{pages}{9747--9754} (\bibinfo{year}{2020}).

\bibitem{andrault1998pressure}
\bibinfo{author}{Andrault, D.}, \bibinfo{author}{Fiquet, G.}, \bibinfo{author}{Guyot, F.} \& \bibinfo{author}{Hanfland, M.}
\newblock \bibinfo{journal}{\bibinfo{title}{Pressure-induced landau-type transition in stishovite}}.
\newblock {\emph{\JournalTitle{Science}}} \textbf{\bibinfo{volume}{282}}, \bibinfo{pages}{720--724} (\bibinfo{year}{1998}).

\bibitem{lichtenberg2021Redox}
\bibinfo{author}{Lichtenberg, T.}
\newblock \bibinfo{journal}{\bibinfo{title}{Redox hysteresis of super-earth exoplanets from magma ocean circulation}}.
\newblock {\emph{\JournalTitle{The Astrophysical Journal Letters}}} \textbf{\bibinfo{volume}{914}}, \bibinfo{pages}{L4} (\bibinfo{year}{2021}).

\bibitem{young2024Phase}
\bibinfo{author}{Young, E.~D.}, \bibinfo{author}{Stixrude, L.}, \bibinfo{author}{Rogers, J.~G.}, \bibinfo{author}{Schlichting, H.~E.} \& \bibinfo{author}{Marcum, S.~P.}
\newblock \bibinfo{title}{Phase equilibria of sub-{{Neptunes}} and super-{{Earths}}} (\bibinfo{year}{2024}).
\newblock \eprint{2408.11321}.

\bibitem{moser2011Pressure}
\bibinfo{author}{Moser, D.} \emph{et~al.}
\newblock \bibinfo{journal}{\bibinfo{title}{The pressure--temperature phase diagram of {{MgH}}{\textsubscript{2}} and isotopic substitution}}.
\newblock {\emph{\JournalTitle{Journal of Physics: Condensed Matter}}} \textbf{\bibinfo{volume}{23}}, \bibinfo{pages}{305403}, \doiprefix\url{10.1088/0953-8984/23/30/305403} (\bibinfo{year}{2011}).

\bibitem{fukai2003Phase}
\bibinfo{author}{Fukai, Y.}, \bibinfo{author}{Mori, K.} \& \bibinfo{author}{Shinomiya, H.}
\newblock \bibinfo{journal}{\bibinfo{title}{The phase diagram and superabundant vacancy formation in {{Fe}}--{{H}} alloys under high hydrogen pressures}}.
\newblock {\emph{\JournalTitle{Journal of Alloys and Compounds}}} \textbf{\bibinfo{volume}{348}}, \bibinfo{pages}{105--109}, \doiprefix\url{10.1016/S0925-8388(02)00806-X} (\bibinfo{year}{2003}).

\bibitem{wang2009High}
\bibinfo{author}{Wang, S.}, \bibinfo{author}{Mao, H.-k.}, \bibinfo{author}{Chen, X.-J.} \& \bibinfo{author}{Mao, W.~L.}
\newblock \bibinfo{journal}{\bibinfo{title}{High pressure chemistry in the {{H}}{\textsubscript{2}}-{{SiH}}{\textsubscript{4}} system}}.
\newblock {\emph{\JournalTitle{Proceedings of the National Academy of Sciences}}} \textbf{\bibinfo{volume}{106}}, \bibinfo{pages}{14763--14767}, \doiprefix\url{10.1073/pnas.0907729106} (\bibinfo{year}{2009}).

\bibitem{strobel2009Novel}
\bibinfo{author}{Strobel, T.~A.}, \bibinfo{author}{Somayazulu, M.} \& \bibinfo{author}{Hemley, R.~J.}
\newblock \bibinfo{journal}{\bibinfo{title}{Novel pressure-induced interactions in silane-hydrogen}}.
\newblock {\emph{\JournalTitle{Physical Review Letters}}} \textbf{\bibinfo{volume}{103}}, \bibinfo{pages}{065701}, \doiprefix\url{10.1103/PhysRevLett.103.065701} (\bibinfo{year}{2009}).

\bibitem{sugimura2008Compression}
\bibinfo{author}{Sugimura, E.} \emph{et~al.}
\newblock \bibinfo{journal}{\bibinfo{title}{Compression of {{H}}{\textsubscript{2}}{{O}} ice to 126 {{GPa}} and implications for hydrogen-bond symmetrization: {{Synchrotron}} x-ray diffraction measurements and density-functional calculations}}.
\newblock {\emph{\JournalTitle{Physical Review B}}} \textbf{\bibinfo{volume}{77}}, \bibinfo{pages}{214103}, \doiprefix\url{10.1103/PhysRevB.77.214103} (\bibinfo{year}{2008}).

\bibitem{degtyareva2007Crystal}
\bibinfo{author}{Degtyareva, O.} \emph{et~al.}
\newblock \bibinfo{journal}{\bibinfo{title}{Crystal structure of {{SiH}}{\textsubscript{4}} at high pressure}}.
\newblock {\emph{\JournalTitle{Physical Review B}}} \textbf{\bibinfo{volume}{76}}, \bibinfo{pages}{064123}, \doiprefix\url{10.1103/PhysRevB.76.064123} (\bibinfo{year}{2007}).

\bibitem{moriwaki2006Structural}
\bibinfo{author}{Moriwaki, T.}, \bibinfo{author}{Akahama, Y.}, \bibinfo{author}{Kawamura, H.}, \bibinfo{author}{Nakano, S.} \& \bibinfo{author}{Takemura, K.}
\newblock \bibinfo{journal}{\bibinfo{title}{Structural phase transition of rutile-type {{MgH}}{\textsubscript{2}} at high pressures}}.
\newblock {\emph{\JournalTitle{Journal of the Physical Society of Japan}}} \textbf{\bibinfo{volume}{75}}, \bibinfo{pages}{074603}, \doiprefix\url{10.1143/JPSJ.75.074603} (\bibinfo{year}{2006}).

\bibitem{cui2008Structural}
\bibinfo{author}{Cui, S.}, \bibinfo{author}{Feng, W.}, \bibinfo{author}{Hu, H.}, \bibinfo{author}{Feng, Z.} \& \bibinfo{author}{Wang, Y.}
\newblock \bibinfo{journal}{\bibinfo{title}{Structural phase transitions in {{MgH}}{\textsubscript{2}} under high pressure}}.
\newblock {\emph{\JournalTitle{Solid State Communications}}} \textbf{\bibinfo{volume}{148}}, \bibinfo{pages}{403--405}, \doiprefix\url{10.1016/j.ssc.2008.09.033} (\bibinfo{year}{2008}).

\bibitem{kippenhahn1991Stellar}
\bibinfo{author}{Kippenhahn, R.} \& \bibinfo{author}{Weigert, A.}
\newblock \emph{\bibinfo{title}{Stellar Structure and Evolution}}.
\newblock Astronomy and Astrophysics Library (\bibinfo{publisher}{Springer-Verl}, \bibinfo{address}{Berlin Heidelberg Paris [etc.]}, \bibinfo{year}{1991}).

\bibitem{turner64DDC}
\bibinfo{author}{{Turner}, J.~S.} \& \bibinfo{author}{{Stommel}, H.}
\newblock \bibinfo{journal}{\bibinfo{title}{{A New Case of Convection in the Presence of Combined Vertical Salinity and Temperature Gradients}}}.
\newblock {\emph{\JournalTitle{Proceedings of the National Academy of Science}}} \textbf{\bibinfo{volume}{52}}, \bibinfo{pages}{49--53}, \doiprefix\url{10.1073/pnas.52.1.49} (\bibinfo{year}{1964}).

\end{thebibliography}

\clearpage

\clearpage

\clearpage

\newpage

\section*{Methods}

\subsection*{Sample Materials and Preparation}
We studied three starting materials: (1) San Carlos olivine, $(\mathrm{Mg}_{0.9}\mathrm{Fe}_{0.1})_2\mathrm{SiO}_4$, (2) natural fayalite, $\mathrm{Fe_{2}SiO_4}$ (Smithsonian: R-3517-00: Rockport, Massachusetts), and (3) silica (Alfa Aesar 99.995\% purity).
For laser coupling, silica was mixed with Fe metal powder (Aldrich 99.9\%+ purity). 
Although San Carlos olivine naturally contains enough Fe for laser coupling, to further improve the coupling, we mixed San Carlos olivine with Fe metal powder (20 wt$\%$). 
Powders were ground and mixed in an alumina mortar, then cold-pressed into foils with approximately 10\,$\mu$m thick. 
Rhenium gaskets were indented by diamond anvils with 200\,$\mu$m diameter culets and then drilled with 125\,$\mu$m diameter holes. 
The rhenium gaskets were then coated with $\sim$800~\AA~of gold to inhibit hydrogen diffusion into the gasket material. 
The gaskets were placed back onto the diamond culets, followed by the sample foils and spacers (of the same material as the sample foil), and gold and ruby grains for pressure calibration. 
The cells were then loaded with 1300--1500\,bar of pure H$_2$ gas in a Sanchez GLS 1500 gas loading system and then compressed to the target pressures (measured using ruby fluorescence \citeM{piermarini1975calibration}) at 300~K before synchrotron laser-heating experiments.
A fayalite sample was loaded with a H$_2$ + Ar gas mixture (50:50) in order to examine the impact of lower concentration of H$_2$.

\subsection*{Synchrotron Experiments}

In-situ X-ray diffraction (XRD) images of the samples in the laser-heated diamond-anvil cell (LHDAC) were collected at the 13-IDD beamline of the GeoSoilEnviroConsortium for Advanced Radiation Sources (GSECARS) sector at the Advanced Photon Source (APS) synchrotron facility. 
Near-infrared laser beams of wavelength 1064 nm and monochromatic X-ray beams of wavelength 0.4133\,{\AA}\ or 0.3344\,{\AA}\ were coaxially aligned and focused on the sample in the LHDAC\@ \citeM{prakapenka2008advanced}. 
Standard continuous laser heating of H-loaded samples in the DAC results in extremely mobile and diffusive hydrogen fluid which can penetrate into diamond anvils, leading to diamond embrittlement and failure of the anvils \citeM{deemyad2005pulsed}.
To enable melting of silicates in a hydrogen medium, a pulsed laser heating system was utilized to mitigate the amount of hydrogen diffusion into the anvils and the gasket material \cite{goncharov2010x}. 
Each pulsed heating event consisted of $10^5$ pulses at 10\,kHz and 20 streak spectroradiometry measurements. 
The pulse width was $1\,\mu$s. 
Therefore, an event with $10^5$ pulses gives a total heating time of 0.1\,s. 
The X-ray spot size is $3{\times}4~\mu\mathrm{m}^2$ and laser-heating spot is an approximately 15~$\mu$m-diameter circle. 
The laser pulses were synchronized with the synchrotron X-ray detector such that diffraction measurements can take place only when the sample reaches the highest temperature during heating. %
The small X-ray beam size and large laser-heating spot size help mitigate the effects from the radial thermal gradients in the high-temperature diffraction patterns. 
Ref.~\citeM{prakapenka2008advanced} showed that the laser heating system provides a flat top laser beam intensity profile which further reduces the radial thermal gradients in the hot spot. 
We also conducted two-dimensional XRD mapping after temperature quench of the samples in LHDAC at high pressures to monitor the possible effects of thermal gradients during laser heating.
In the \textit{in situ} high-$T$ XRD patterns, it was difficult to clearly distinguish broad diffuse scattering features from melt due to the short duration of the measurements.

Although the heating duration is short, it was found that hydrogen is extremely reactive with molten silicates at high temperatures, overcoming the limited heating exposure time. 
In addition, sample foils cold-compressed from powder permit hydrogen gas and fluid to percolate and surround individual grains (less than 1~$\mu$m in size). 
This creates a very large surface area of the silicate exposed to hydrogen and therefore facilitates a fast reaction \citeM{fu2022stable}.
A recent study obtained consistent results between short pulse heating and continuous heating under hydrogen-rich conditions in diamond-anvil cells \citeM{fu2023CoreOriginSeismic}. 

Thermal emission spectra from both sides of the sample in LHDAC were fitted to the gray-body equation to estimate the temperatures \citeM{prakapenka2008advanced}. 
Temperatures are assigned from an average of temperatures recorded from 20 measurements (10 each up and down stream) over a heating event (Extended Data Tab.~\ref{tab:SI:run-table}).
Temperature uncertainty was calculated from the standard deviations $(1\sigma)$ of these measurements.
If the standard deviation is smaller than 100~K, from intrinsic uncertainties in the spectroradiometry method we assigned 100~K for the uncertainty \citeM{kulka2020Bridgmanite}.

A Dectris Pilatus 1M CdTe detector was used to collect two-dimensional diffraction images.
The diffraction images are integrated to one-dimensional diffraction patterns using the DIOPTAS package \citeM{prescher2015dioptas}. 
Diffraction patterns of CeO$_{2}$ and LaB$_{6}$ were measured for the correction of detector tilt and the determination of the sample-to-detector distance. 
Unit-cell parameter fitting was conducted by fitting the diffraction peaks with pseudo-Voigt profile functions in the PeakPo package \citeM{PeakPo}. 
Pressure was calculated from the unit-cell volume of a gold grain at the edge of the sample chamber using the equation of state of gold \citeM{ye2017intercomparison} before and after heating.
A gold grain was placed away from the sample rather than mixed with it to prevent reactions/alloying with the sample material, and thus pressure could not be measured at high temperatures during laser heating.  %
A previous study \citeM{dewaele1998temperature} showed that the thermal pressure in a liquid Ar medium at temperatures of 1000--4000~K is 0.5--2.5~GPa in LHDAC\@. 
All of our experiments exceed the melting temperature of hydrogen.
Therefore, thermal pressure should be similar to the above estimation in our experiments, and we assign a pressure uncertainty of 10\% during laser heating \cite{horn2023reaction}. 
We note that this method does not introduce a severe error for the purpose of this study, which is to explore hydrogen-silicate reactions at high pressures.

\subsection*{Raman Spectroscopy}

Raman measurements were conducted utilizing the Raman spectroscopy system at GSECARS \citeM{holtgrewe2019advanced} for the identification of O-H and Si-H vibrations after heating. 
Raman scattering of the sample in a diamond-anvil cell was excited by a monochromatic 532-nm beam from a Coherent VERDI V2 laser. 
Raman spectra were collected over a wide range of wavenumbers (1400--4500~cm$^{-1}$) utilizing a Princeton Instruments Acton Series SP-2560 spectrograph and PIXIS100 detector.

\subsection*{Modeling for the Dynamics of the Interior}

The thermal evolution model is calculated for the entire interior (from center to surface) on one mass grid, with no distinction between core and envelope (Fig.~\ref{fig:AV2}). 
The model is 1D and solves the standard interior structure and evolution equations, which allow for heat transport by convection, radiation, and conduction, depending on local conditions over time (equations can be found in ref.~\citeM{vazan2015convection}). 
The basic set of parameters for rocky planets with gas envelopes is adopted from ref.~\cite{vazan2018Contribution}. 
The redistribution of composition by convective mixing where the convection criterion is fulfilled, is calculated according to ref.~\citeM{vazan2015convection}. 

The input equations of states are of ref.~\citeM{saumon1995equation} for hydrogen and helium, and an improved version of ref.~\citeM{vazan2013effect} for water and silica as representatives of volatiles and refractories, respectively. 
In our experiments, we found that multiples of components, such as SiH$_4$ and MgH$_2$, in addition to H$_2$O can exist in H$_2$ envelopes. In such a system semi convection (double diffusive convection) might develop under certain conditions (Supplementary Discussion 14), which could limit the mixing efficiency and terminate water production earlier than appears in our models.
Irradiation by the parent star is included as a temperature boundary for a plane-parallel gray atmosphere with an optical depth of 1. 
The radiative opacity is that of a grain-free solar metallicity atmosphere \citeM{freedman2014gaseous}. 

\section*{Data availability}

X-ray diffraction and Raman data can be found in Zenodo with the identifier 10.5281/zenodo.15586691 \citeM{shim2025Jupyter}.  
An overview of the data is also included there.

\section*{Materials \& Correspondence}
Correspondence and requests for additional information should be addressed to H.H. (horn24@llnl.gov, hallensu@asu.edu) or S.H.S. (sshim5@asu.edu).

\clearpage
\bibliographystyleM{naturemag-doi}
\bibliographyM{dis,add}

\newpage
\section*{Acknowledgments}
We thank the three anonymous reviewers for their valuable comments and insightful questions, which significantly improved the quality of this paper.
S.-H.S and H.W.H were supported by National Science Foundation (NSF) Grants AST-2108129, AST-2406790, and EAR-1921298.
Portions of this work were performed at GeoSoilEnviroCARS (The University of Chicago, Sector 13), Advanced Photon Source (APS), Argonne National Laboratory. 
GeoSoilEnviroCARS is supported by the National Science Foundation - Earth Sciences (EAR-1634415) and Department of Energy (DOE) - GeoSciences (DE-FG02-94ER14466). 
This research used resources of the Advanced Photon Source, a U.S. DOE Office of Science User Facility operated for the DOE Office of Science by Argonne National Laboratory under Contract No. DE-AC02-06CH11357. 
We acknowledge the use of facilities within the Eyring Materials Center at ASU\@. 
Part of this work was performed under the auspices of the U.S. Department of Energy by Lawrence Livermore National Laboratory under Contract DE-AC52-07NA27344. 
The opinions are those of the author and do not necessarily represent the opinions of LLNL, LLNS, DOE, NNSA or the US government.
The experimental data for this paper are available by contacting hallensu@asu.edu. 
A.V. acknowledges support by ISF grants 770/21 and 773/21.

\section*{Author contributions statement}
H.H. and S.H.S. conceived the project, H.H., S.C., V.B.P., and S.H.S. conducted synchrotron experiments, A.V. conducted sub-Neptune modeling, H.H. and S.H.S. analyzed the results, H.H., S.H.S., and A.V. wrote the manuscript.
All authors reviewed the manuscript.

\section*{Additional information}

The authors declare no conflict of interests. This manuscript has been authored by Lawrence Livermore National Security, LLC under Contract No. DE-AC52-07NA27344 with the US. Department of Energy. The United States Government retains, and the publisher, by accepting the article for publication, acknowledges that the United States Government retains a non-exclusive, paid-up, irrevocable, world-wide license to publish or reproduce the published form of this manuscript, or allow others to do so, for United States Government purposes. Release authorization LLNL-JRNL-872168.

\newpage
\setcounter{equation}{0}
\setcounter{figure}{0}
\setcounter{table}{0}
\makeatletter
\renewcommand{\figurename}{\bf Extended Data Fig.}
\renewcommand{\tablename}{\bf Extended Data Table}
\renewcommand{\theequation}{Extended Data Reaction \arabic{equation}}
\newpage

\strutlongstacks{T}
\begin{table}[]
\caption{Experimental runs in this study. 
Estimated uncertainties for pressure $(P)$ is 10\% (see the Method section).
In some runs, multiples of heating events ($10^5$ laser pulses at 10~kHz) were performed and therefore temperature for each individual heating events are provided.
Fe$_{1-y}$Si$_y$ was observed in the B2 structure throughout these runs.
FeH$_x$ appears in the face-centered cubic (fcc) structure during heating and temperature-quenching.  
When Fe metal was loaded as a starting material (SCO and SIL runs), double hexagonal close packed (dhcp) structured FeH$_x$ was also observed.
Fe metal (Fe) was observed as body-centered cubic (bcc) at pressures lower than 10~GPa in some FAY runs.
A.M.: analytical methods; XRD: X-ray diffraction; SEM: scanning electron microscopy; Raman: Raman spectroscopy;
Rwd: ringwoodite; Wds: wadsleyite; Bdm: bridgmanite; Sti: stishovite; Coe: coesite.
}
\resizebox{0.7\textheight}{!}{
\begin{tabular}{lclll}
\hline
Run & $P$ (GPa) & $T$ (K) & A.M. & Result \\\hline

\multicolumn{5}{c}{(Mg$_{0.9}$Fe$_{0.1}$)$_2$SiO$_4$ + Fe + H} \\

SCO-1 & 8 & 3042(159) & XRD, SEM & Fe$_{1-y}$Si$_y$, MgO \\

SCO-2 & 6 & 2833(177) & XRD, SEM & FeH$_x$, MgO  \\ %

SCO-3 & 8 & 3104(269) & XRD, SEM & Fe$_{1-y}$Si$_y$, FeH$_x$, MgO \\

SCO-4 & 7 & 3050(100) & XRD, SEM & Fe$_{1-y}$Si$_y$, FeH$_x$, MgO \\

SCO-5 & 7 & 2777(100) & XRD, SEM & Fe$_{1-y}$Si$_y$, FeH$_x$, MgO  \\

SCO-6 & 17 & 3074(267) & XRD & \Shortunderstack[l]{Fe$_5$Si$_3$,~Fe$_{1-y}$Si$_y$,~FeH$_x$,\\ \hspace{.2cm} MgO,~Rwd,~Wds} \\

SCO-7 & 21 & 2534(218) & XRD & \Shortunderstack[l]{Fe$_5$Si$_3$,~FeH$_x$,~Sti,\\ \hspace{.2cm} MgO,~Rwd} \\

SCO-8 & 21 & 2738(119), 2989(269) & XRD & \Shortunderstack[l]{Fe$_5$Si$_3$,~FeH$_x$,~Sti,~MgO} \\

SCO-9 & 22 & 3101(462) & XRD & \Shortunderstack[l]{Fe$_5$Si$_3$,~Fe$_{1-y}$Si$_y$,~FeH$_x$,\\ \hspace{.2cm} Sti,~MgO,~Rwd} \\

SCO-10 & 30 & \Shortunderstack[l]{3298(150), 2434(100), 2264(100),\\ 2187(100), 2310(100), 2539(100), \\ 4035(144)} & XRD, SEM & \Shortunderstack{Fe$_{1-y}$Si$_y$,~FeH$_x$,~MgO,~Bdm} \\

SCO-11 & 37 & 3557(677) & XRD & Fe$_{1-y}$Si$_y$, FeH$_x$, MgO, Bdm  \\

SCO-12 & 42 & 2704(150), 2467(150) & XRD, SEM & FeH$_x$, Bdm, MgO \\

SCO-13a & 42 & \Shortunderstack[l]{2792(162), 2725(133), 3075(150),\\ 3197(269)} & XRD, SEM & FeH$_x$, Bdm, MgO \\

SCO-13b & 42 & \Shortunderstack[l]{3789(378), 3352(174), 3924(258)} &  XRD, SEM & Fe$_{1-y}$Si$_y$, FeH$_x$, MgO \\

\multicolumn{5}{c}{Fe$_2$SiO$_4$ + H} \\

FAY-1 & 6 & 3017(402) & XRD, Raman, SEM & Fe$_{1-y}$Si$_y$, FeH$_x$, Fe \\

FAY-2 & 6 & 2254(100) & XRD, Raman & Fe$_{1-y}$Si$_y$, FeH$_x$, Fe \\

FAY-3 & 11 & 2898(427) & XRD, Raman, SEM & FeH$_x$, Sti, Coe\\ %

FAY-4 & 11 & 2808(475), 2584(166) & XRD, Raman & FeH$_x$, Sti, Coe\\

FAY-5 & 21 & 3965(407), 3715(954) & XRD, Raman & FeH$_x$, Sti \\

\multicolumn{5}{c}{Fe$_2$SiO$_4$ + H + Ar (50\% H$_2$)} \\

FAY-6 & 10 & 3500(350) & XRD, Raman & FeH$_x$, Sti, Coe\\

FAY-7 & 10 & 1100(500) & XRD, Raman & FeH$_x$, Sti, Coe\\

FAY-8 & 14 & 1550(150) & XRD, Raman & FeH$_x$, Sti, Coe\\

\multicolumn{5}{c}{SiO$_2$ + Fe + H} \\

SIL-1 & 14 & 1934(526) & XRD, Raman & Fe$_{1-y}$Si$_y$, FeH$_x$, Sti, Coe \\

SIL-2 & 14 & 2899(282) & XRD, Raman & Fe$_{1-y}$Si$_y$, Sti, Coe\\

SIL-3 & 14 & 2750(100) & XRD, Raman & Fe$_{1-y}$Si$_y$, FeH$_x$, Sti \\ %

SIL-4 & 39 & 2149(514) & XRD & Fe$_{1-y}$Si$_y$, Sti \\

SIL-5 & 39 & \Shortunderstack[l]{1647(258), 2906(254), 3039(284),\\ 3009(100), 3103(100), 3232(206)} & XRD & Fe$_{1-y}$Si$_y$, Sti \\\hline

\end{tabular}}\label{tab:SI:run-table}
\label{table}
\end{table}

\clearpage
\begin{figure}[!ht]
\centering
\includegraphics[width=0.45\textwidth]{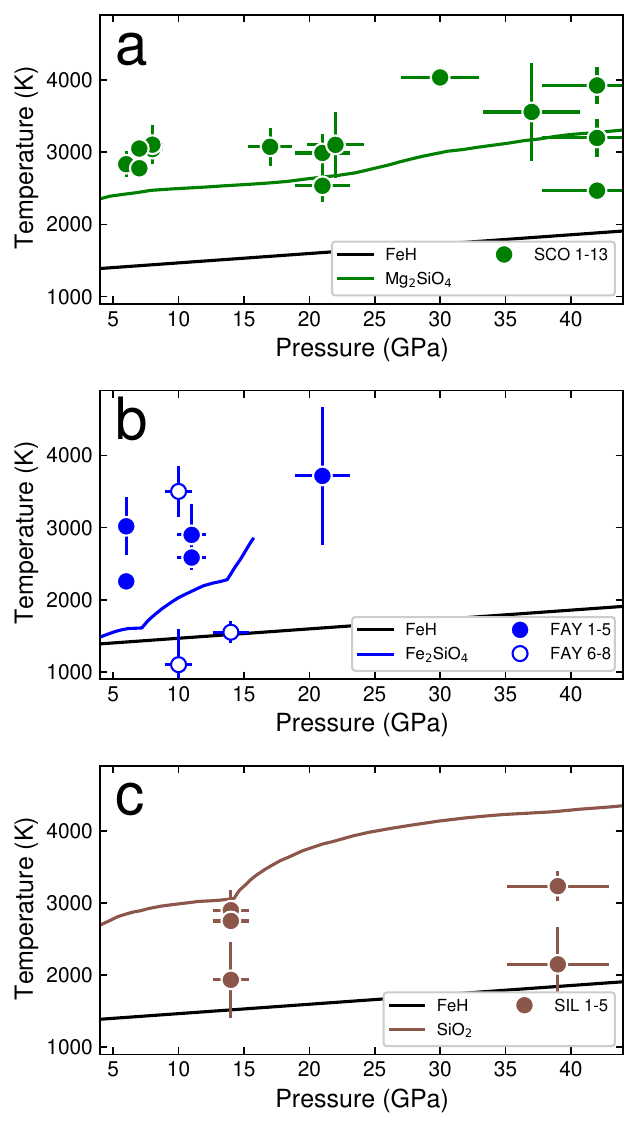}
\caption{Pressure-temperature conditions of the experimental runs from \textbf{a,} San Carlos olivine, \textbf{b,} fayalite, and \textbf{c,} silica starting materials in this study (Extended Data Table~\ref{tab:SI:run-table}).
For fayalite, three experimental runs were conducted with a 50\% Ar + 50\% H$_2$ medium (open circles).
Melting curves for the relevant phases are shown: FeH$_x$ (ref.~\protect\citeM{sakamaki2009}); Mg$_{2}$SiO$_{4}$ (ref.~\protect\citeM{mosenfelder2007thermodynamic}); Fe$_{2}$SiO$_{4}$ (ref.~\protect\citeM{ohtani1979melting}); SiO$_2$ (ref.~\protect\citeM{andrault2020melting}).
All the experiments were conducted above the melting temperature of hydrogen \protect\citeM{zha2017Melting}. 
While olivine and fayalite couple with laser beams sufficiently well for melting, lack of Fe in silica makes it difficult to heat above melting as shown in {\bf c}.
}\label{fig:SI:PT}%
\end{figure}

\clearpage

\begin{figure}[!ht]
    \centering
    \includegraphics[width=0.6\textwidth]{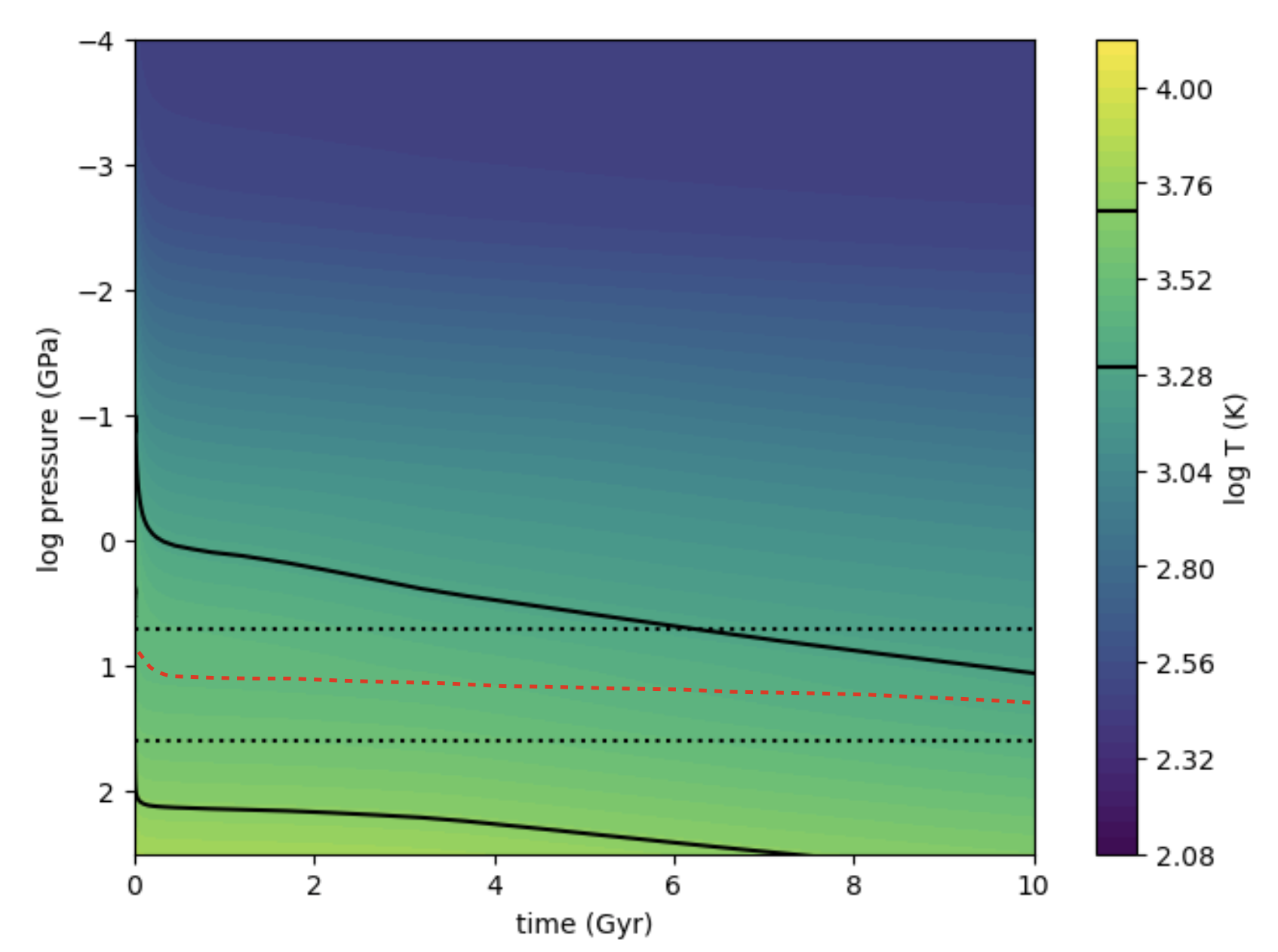}
    \caption{Thermal evolution of a rocky $5\,M_{E}$ sub-Neptune planet with an H + He (5\%) envelope. 
    Shown is the temperature (color) as a function of pressure ($y$-axis) in the interior from the center up to 1~bar pressure, as a function of time ($x$-axis). 
    The range of pressure (dotted black) and temperature (solid black) in which water production is expected according to the experiments is shown. 
    The red dashed line signifies the mantle-envelope interface. 
    Model is based on ref.~\protect\cite{vazan2018Contribution}.}\label{fig:SI:amb-PT} %
\end{figure}

\clearpage
\begin{figure}[!ht]
\centering
\includegraphics[width=1\textwidth]{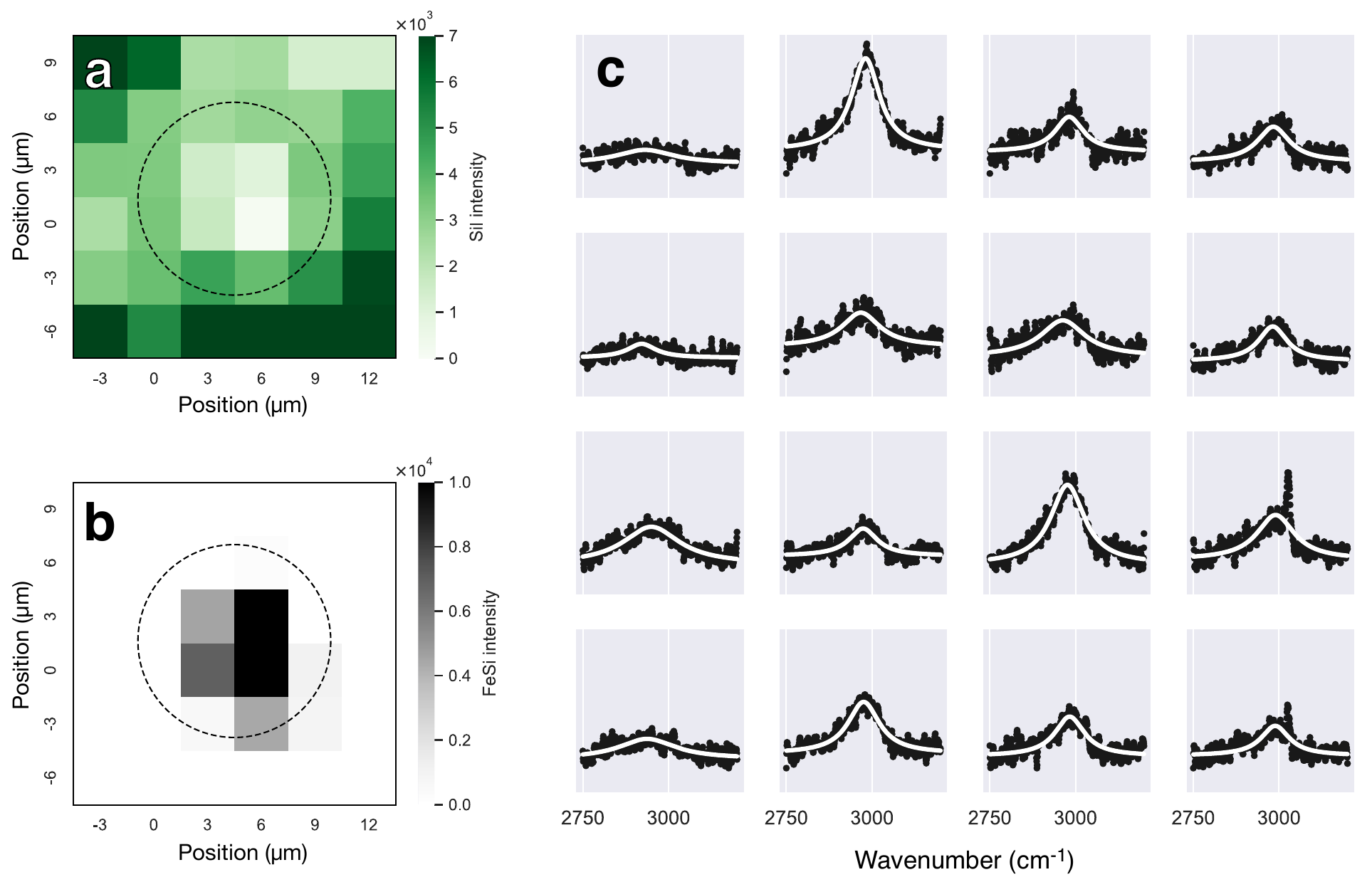}
\caption{{\bf Analysis of the samples after heating in a hydrogen medium in LHDAC\@.}
Two dimensional maps of the XRD intensities of {\bf a,} silicates and {\bf b,} B2~$\mathrm{Fe}_{1-y}\mathrm{Si}_y$ after heating the sample in SCO-11. 
The anti-correlation between the two at the heating spot center (dashed circle) shows that when melted silicates (bdm) break down, Si is reduced to form $\mathrm{Fe}_{1-y}\mathrm{Si}_y$.
{\bf c,} Raman-active OH vibration from H$_2$O ice after heating silica + Fe metal.
The Raman spectra were measured for a $20{\times}20\,\mu\mathrm{m}^{2}$ heated area. 
The distance between the spots where the spectra were collected is 5~$\mu$m.
The spectra were collected after laser heating at 14~GPa.
}\label{fig:SI:map}
\end{figure}

\clearpage
\begin{figure}[htbp!]
\centering
\includegraphics[width=0.5\textwidth]{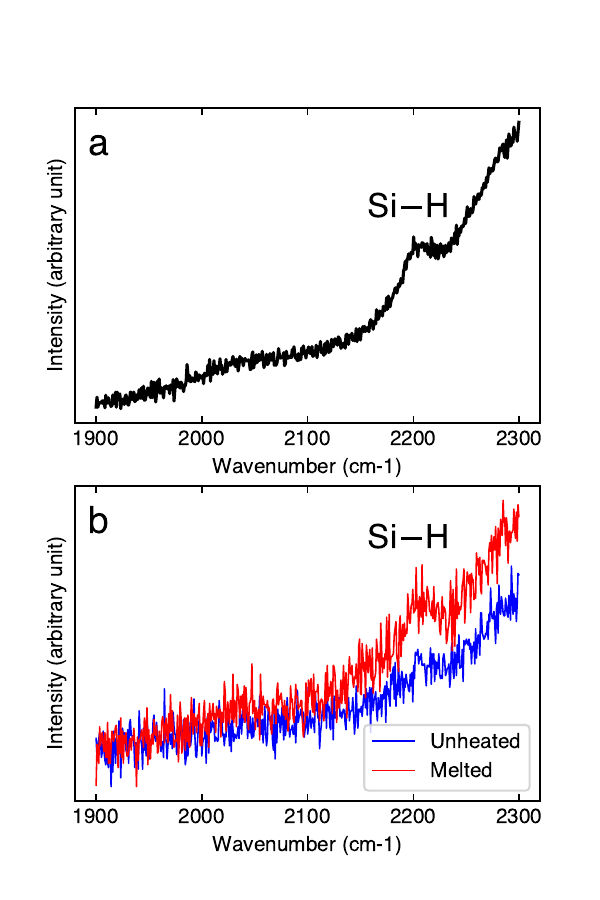}
\caption{Raman spectra from run FAY-8 where starting fayalite was heated to 1550~K at 14~GPa.
The measurements were conducted after temperature quench to 300~K and decompression to 2.5~GPa\@.
The Si--H vibrational mode was detected at the melted area (\textbf{a}). 
No such feature was observed outside the melted area (blue, \textbf{b}).
In \textbf{b}, we also include spectrum measured at the melted spot (red) for the same exposure time.
The same mode has been documented for silica melted in hydrogen at 2--3~GPa (ref.~\cite{shinozaki2014formation}).
}\label{fig:SI:SiH4}
\end{figure}

\clearpage
\begin{figure}[htbp!]
\centering
\includegraphics[width=0.8\textwidth]{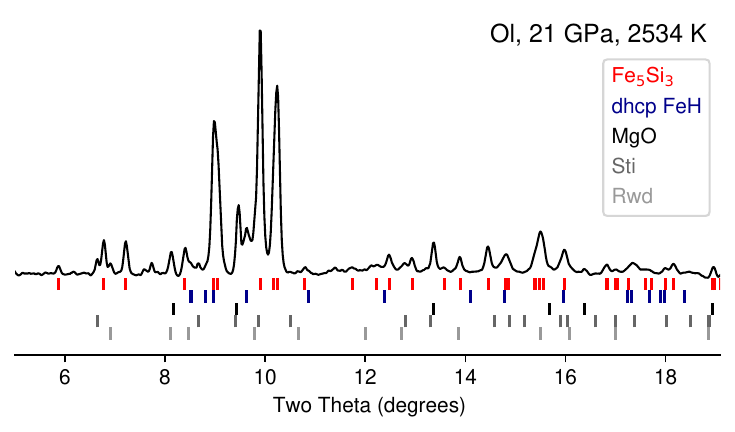}
\caption{X-ray diffraction pattern measured after heating a starting mixture of San Carlos olivine and Fe metal to 2534~K at 21~GPa (run SCO-7).
The ticks below the diffraction pattern are the peak positions of the observed phases.
The names of the phases in the legend is ordered same as the ticks from top to bottom.
X-ray energy is 37~keV\@.
}\label{fig:SI:xrd_olv21}
\end{figure}

\clearpage

\begin{figure}[htbp!]
\centering
\includegraphics[width=0.8\textwidth]{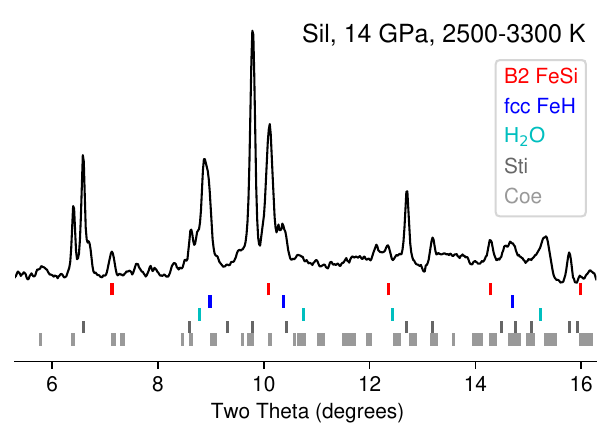}
\caption{X-ray diffraction pattern measured after heating a silica starting material in a hydrogen medium to 2899~K (below the melting temperature) at 14~GPa (run SIL-2).
The ticks below the diffraction pattern are the peak positions of the observed phases.
The names of the phases in the legend is ordered same as the ticks from top to bottom.
X-ray energy is 37~keV\@.
}\label{fig:SI:SiO2}
\end{figure}

\clearpage
\begin{figure}[htbp!]
\centering
\includegraphics[width=\textwidth]{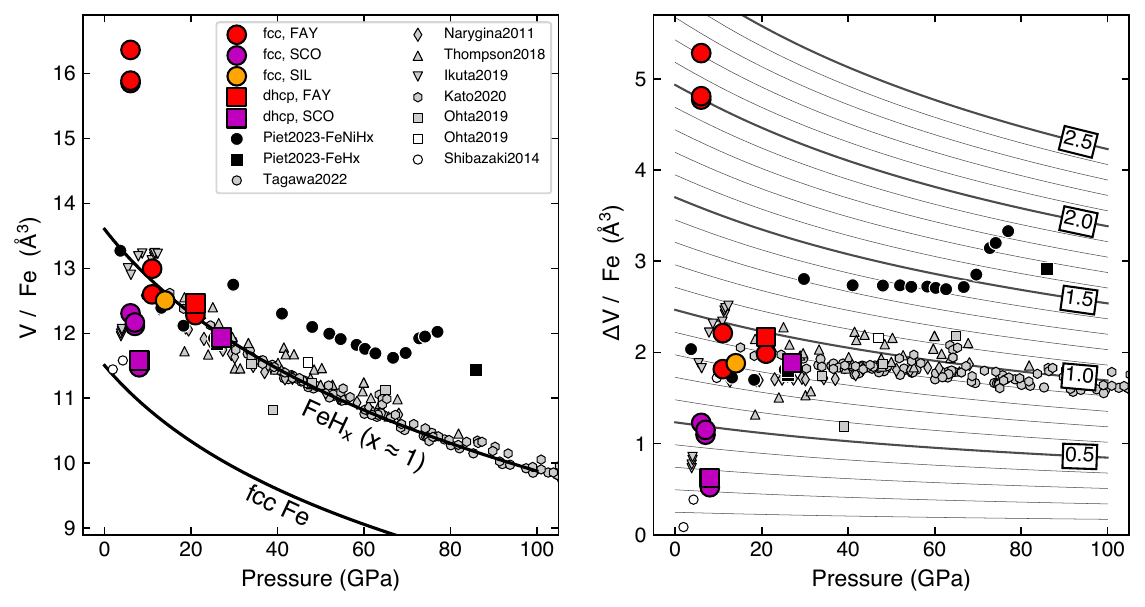}
\caption{Atomic volume (left) and volume increase by H incorporation, $\Delta V$ (right), of FeH$_x$ observed in different runs (colored symbols). 
The volumes were measured after heating at 300~K\@.
For comparison, the figures also include data points from previous studies \protect\citeM{narygina2011x,thompson2018high,kato2020hydride,tagawa2022HighTemperature,ikuta2019interstitial,shibazaki2014high,ohta2019electrical}.
The equations of state for fcc~Fe (H/Fe = 0) and fcc~FeH (H/Fe = 1) are from ref.~\protect\citeM{dorogokupets2017Thermodynamics} and ref.~\protect\citeM{kato2020hydride}, respectively.
The data points measured for the solid phases quenched from (Fe,Ni)-H liquid are shown as black symbols \protect\citeM{piet2023}.
The concentration curves $(x)$ shown in the right figure are from the density functional theory calculation in ref.~\protect\citeM{piet2023}.
}\label{fig:SI:FeH-vol}%
\end{figure}

\clearpage
\begin{figure}[!ht]
\centering
\includegraphics[width=0.7\textwidth]{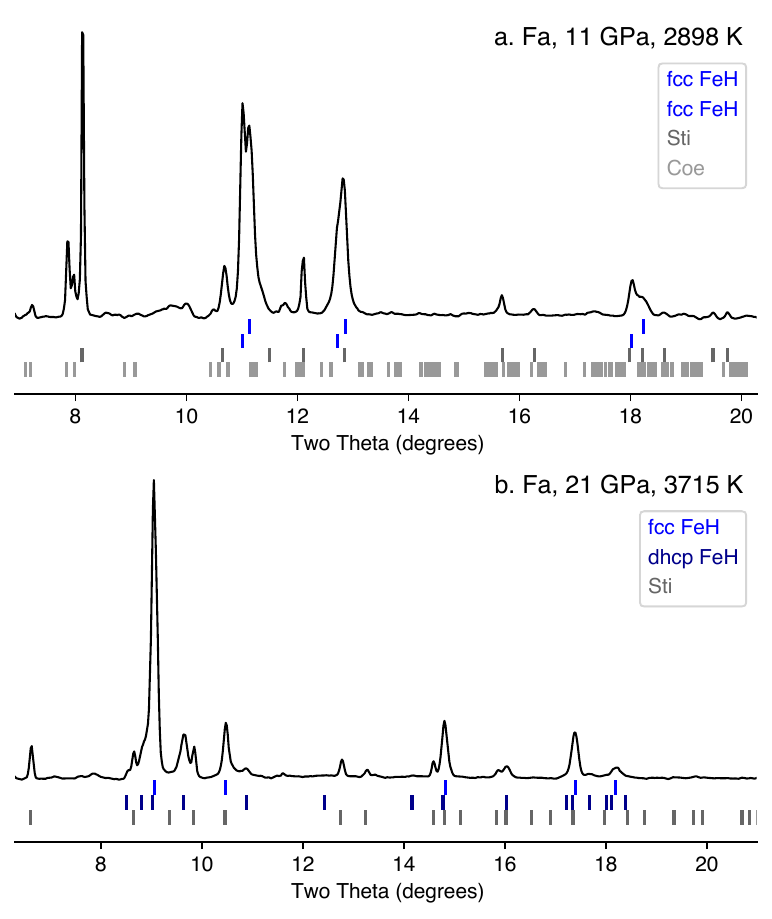}
\caption{X-ray diffraction pattern measured after heating a fayalite starting material in a hydrogen medium to {\bf a,} 2898~K at 11~GPa (run FAY-3) and {\bf b,} 3715~K at 21~GPa (run FAY-5).
{\bf a} was measured at 6~$\mu$m away from the heating center. 
Two separate fcc phases with different volumes (and therefore different levels of hydrogenation) are observed, likely because of different rate of temperature decrease during quenching at different spots and loss of hydrogen during quench of FeH$_x$ melt \protect\citeM{piet2023}.
The ticks below the diffraction patterns are the peak positions of the observed phases.
The names of the phases in the legend is ordered same as the ticks from top to bottom.
X-ray energy is {\bf a,} 30~keV and {\bf b,} 37~keV.}
\label{fig:SI:fay21}
\end{figure}

\clearpage
\begin{figure}[!ht]
\centering
\includegraphics[width=0.8\textwidth]{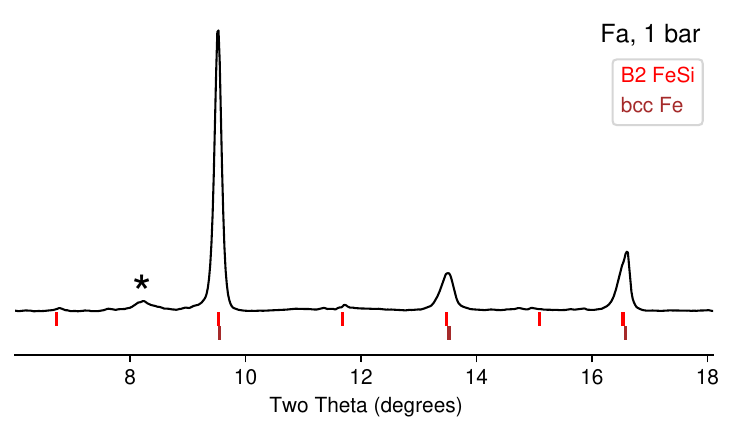}
\caption{X-ray diffraction pattern measured after the decompression of a fayalite starting material heated in a hydrogen medium to 3017~K at 6~GPa (run FAY-1).
The diffraction pattern was measured at 1~bar and 300~K\@.
The ticks below the diffraction pattern are the peak positions of the observed phases.
The names of the phases in the legend is ordered same as the ticks from top to bottom.
``*'' indicates a feature from detector defects.
X-ray energy is 37~keV\@.
}\label{fig:SI:fay1}
\end{figure}

\newpage

\part*{Supplementary Information}
\setcounter{section}{0}

\section{Diamond embrittlement by hydrogen in laser-heated diamond-anvil cell experiments}\label{SI:section:diamond-embrittlement}

In high-$P$ experiments, it has been extremely challenging to melt both hydrogen and silicates together (${>}2000$~K).
As the smallest atom, hydrogen can diffuse into the crystal structures of experimental assembly materials (including diamond), causing embrittlement and ultimately leading to experimental failure\citeS{ji2019ultrahigh}.
Such embrittlement of anvil and gasket materials makes it impossible to investigate essential chemical interactions to understand volatile-rich planets.  

\section{SCO-1, 2, 3, 4, and 5}\label{SI:section:olivine-composition}

\paragraph{Fe$_{1-y}$Si$_y$} 
All Fe-Si alloy that was observed was in the B2 structure and the B20 structure was not observed in this pressure range, which is different from the behaviors observed in H-free Fe-Si alloy \citeS{fischer2014equations}.
A recent study found that the B2 structure can be further stabilized over B20 by hydrogen \citeS{fu2022fesiH}.
In run~SCO-3, the unit-cell volume of the B2 phase is 22.30~\AA$^3$ at 8~GPa,  $\sim$8.5\% larger than the expected volume from the equation of state of B2~FeSi with Si/Fe = 1 (or $y=0.5$) (ref.~\citeS{fischer2014equations}). 
Similar trends were also observed in other SCO runs in this pressure range.
This observation could either be due to hydrogen incorporation or a lower concentration of Si (ref.~\citeS{slater1964atomic}). 
This expansion persists through decompression to 1~bar suggesting that the expansion is likely not due to hydrogen incorporation in Fe metal, as it is known to dehydrogenate below 3.5~GPa (ref.~\citeS{badding1991}). 
This is in agreement with ref.~\citeS{fu2022fesiH} who found that B2~FeSi in a pure H medium does not take on a significant amount of H like Fe metal does. 
The atomic volumes ($V/Z$, volume per atom in the unit cell) at 8~GPa and 1~bar are both in close agreement with that of the DO$_3$ phase of $\mathrm{Fe_{0.73}Si_{0.27}}$ (ref.~\citeS{fischer2014equations}), indicating this phase has a composition similar to this phase. 
At 1~bar after decompression, we still observed B2 Fe-Si alloy alongside dehydrogenated bcc~Fe metal. 

\paragraph{FeH$_x$} 
At this pressure range, FeH$_x$ phases were observed in all the runs but SCO-2.
Both the fcc and dhcp structures were observed.
In run~SCO-3 at 8~GPa, the observed unit-cell volume of fcc~FeH$_x$, 45.91~\AA$^3$, was higher than the expected value for Fe metal without H, 42.94~\AA$^3$, but smaller than that of FeH, 56.42~\AA$^3$ (ref.~\citeS{narygina2011}).
In order to calculate the content of hydrogen in FeH$_x$, we used the method developed in ref.~\protect\citeM{piet2023} whereby the volume expansion by hydrogen incorporation was calculated at different pressures using the equations of state of FeH$_x$ from density functional theory calculations.  
The method yields FeH$_{0.22}$ for the observed volume (Extended Data Fig.~\ref{fig:SI:FeH-vol}). %
In the same run, the dhcp structured FeH$_x$ has a unit-cell volume of 46.32~\AA$^3$ compared to the expected unit-cell volume of dhcp~FeH$_x$ with $x \approx 1$ of $\sim$52.5~\AA$^3$ (ref.~\citeS{hirao2004compression}).
However, its volume is expanded by 9.6\% compared to an estimate for hypothetical H-free Fe dhcp structure \citeS{Jain2013}, again indicating substoichiometric alloying ($x\approx 0.4$). 
Upon decompression, FeH$_x$ alloys are known to convert to H-free bcc Fe (ref.~\citeS{badding1991}) and we observed the same. 
The bcc iron shows a small unit-cell parameter of 2.843$\pm$0.003~\AA\ compared to the expected value for pure bcc Fe metal, 2.867~\AA\ (ref.~\citeS{rotter1966ultrasonic}). 
The smaller unit-cell parameter can be explained by a small amount of Si in the quenched bcc phase.

\paragraph{MgO}
MgO was observed in all the SCO runs in this pressure range.
Upon decompression to 1~bar, the unit-cell volume of MgO was 74.66$\pm$0.07~\AA$^3$ in run~SCO-3, in line with the known value of 74.71~\AA$^3$ of pure MgO (ref.~\citeS{utsumi1998volume}), indicating no Fe$^{2+}$ is present in the structure after heating, consistent with reduction of all Fe$^{2+}$ in the starting materials. 
The result is also consistent with a recent experiment on (Mg,Fe)O in a hydrogen medium \cite{horn2023reaction}.

\section{SCO-6, 7, 8, and 9}\label{SI:section:ringwoodite-synthesis}

\paragraph{Fe$_{1-y}$Si$_y$} 
We observed Fe$_5$Si$_3$ (or $y=0.38$) in all the runs in this pressure range.
B2~FeSi was also observed in runs SCO-6 and SCO-9.
Fe$_5$Si$_3$ is known to decompose to FeSi and Fe$_3$Si above 18~GPa and 1300~K in the hydrogen free system \citeS{mcguire2017isothermal}.
However, this phase was observed in a H-bearing medium in a recent experiment~\citeM{fu2022stable}.
Unlike ref.~\citeM{fu2022stable} where hydrogenation of the hexagonal Fe$_5$Si$_3$ phase in a H medium results in significant volume expansion, the unit-cell volume of the Fe$_5$Si$_3$ phase in our study is very close to that expected for H-free Fe$_5$Si$_3$ (ref.~\citeS{errandonea2008structural}), implying minimal hydrogen incorporation into this phase in our experiments. %

\paragraph{FeH$_x$}
As in the other experiments. FeH$_x$ is observed in the fcc structure after high-temperature heating and the dhcp structure before heating or away from the heating center. 
The diffraction intensity is weak compared to that of Fe$_5$Si$_3$ and the complex diffraction pattern (Extended Data Fig.~\ref{fig:SI:xrd_olv21}) precludes the use of the accurate fitting and refinement of the unit-cell parameters of the FeH$_x$ phases. 
However, the observed peak positions are consistent with those expected for $x\approx1$ in FeH$_x$ for both the dhcp and the fcc phases.

\paragraph{MgO} MgO was observed throughout the runs in this pressure range, indicating breakdown of silicates upon melting in a hydrogen medium. 

\paragraph{Ringwoodite/Stishovite} 
Off from the hotspot center, weak peaks of ringwoodite were observed after heating (Extended Data Fig.~\ref{fig:SI:xrd_olv21}).
These diffraction peaks are much weaker than those of $\mathrm{Fe_5Si_3}$ and MgO, suggesting a very small amount of silicate remains.
Often at the center of heated spot, because of the melting of the sample and subsequent reaction, not much starting material remained but instead the area was filled with the hydrogen medium.  
Therefore, high-quality diffraction patterns are normally measured at spots away from the heating center after temperature quench. 
The temperature during heating at the spot where the diffraction patterns showing ringwoodite were measured was lower than the temperature at the heating center and was likely not high enough to achieve melting.
The lack of melting may lead to the incomplete decomposition of the silicate by hydrogen, which is much more significant when silicate is molten.
Stishovite was observed in SCO-7, 8, and 9.
However, as shown in Extended Data Fig.~\ref{fig:SI:xrd_olv21}, their diffraction intensities are low and therefore the amount in the samples is small.

\section{SCO-10, 11, 12, 13a, and 13b}\label{SI:section:bridgmanite-recovered-composition}

\paragraph{Fe$_{1-y}$Si$_y$} 
B2~Fe$_{1-y}$Si$_y$ was observed in runs SCO-10, 11, and 13b where temperature was sufficiently high to melt the silicate.
The volume of B2~Fe$_{1-y}$Si$_y$ is close to the volumes reported for the same phase for Si/Fe = 1 (or $y=0.5$) (ref.~\citeS{fischer2014equations}) at high pressure and through decompression, suggesting little incorporation of H, consistent with the observations in ref.~\citeS{fu2022fesiH} that this phase cannot take in a large amount of H\@. 
At 1~bar, the unit-cell volume of B2~Fe$_{1-y}$Si$_y$ is 21.45$\pm$0.02~\AA$^3$, slightly expanded from the value of 21.30~\AA$^3$ for $y=0.5$ reported by ref.~\citeS{fischer2014equations} (run~SCO-13b). 
This is likely because there is a slight superabundance of Fe (Fe$_{1-y}$Si$_{y}$, $0.33< y < 0.5$).

\paragraph{FeH$_x$} 
We observed FeH$_x$ in all the runs with a volume corresponding to $x \approx 1$.
After pressure quench to 1~bar, FeH$_x$ converts to bcc Fe (run~SCO-13b).
Bcc~Fe has a unit-cell parameter of 2.8666$\pm$0.0001~\AA$^3$ in agreement with the known value of pure Fe metal, 2.867~\AA\ (ref.~\citeS{rotter1966ultrasonic}), suggesting no Si incorporation in FeH$_x$ at this pressure range. 

\paragraph{Ferropericlase and bridgmanite} 
At 1~bar after decompression of the heated samples, ferropericlase has a unit-cell volume of 74.70$\pm$0.02~\AA$^3$, in agreement with the value of 74.71~\AA$^3$ for endmember MgO (i.e., periclase) \citeS{fei1994situ} (run~SCO-13b). 
By contrast, the unit-cell volume of the remnant perovskite phase is 163.01$\pm$0.05\AA$^3$, expanded from the unit-cell volume of endmember MgSiO$_3$ perovskite, 162.49\AA$^3$ (ref.~\citeS{mao1991effect}) (run~SCO-13a). 
The volume is marginally larger than the expected volume for (Mg$_{0.9}$Fe$_{0.1}$)SiO$_3$ of 162.79~\AA$^3$. 
Utilizing a linear interpolation between the unit-cell volumes for 10 and 20~mol\% Fe from ref.~\citeS{mao1991effect} gives an approximate composition of (Mg$_{0.86}$Fe$_{0.14}$)SiO$_3$. 
Note that the bridgmanite was observed in an area away from the center heated to lower temperatures below melting of the starting materials.
When temperature was sufficiently high for melting of silicate (SCO-10, 11, and 13b), bridgmanite diffraction lines are either absent or weak, suggesting that most Si$^{4+}$ was reduced to Si$^0$ or released as SiH$_4$.

\section{Summary on the SIL and FAY runs}\label{SI:section:SiO2-fay-summary}

Pure SiO$_2$ mixed with Fe metal was also heated in a hydrogen medium.  
At 14~GPa after one heating event at 2899~K (SIL-2), SiO$_2$ partially breaks down to alloy with the mixed Fe and form B2~$\mathrm{Fe}_{1-y}\mathrm{Si}_y$ (Extended Data Fig.~\ref{fig:SI:SiO2}), consistent with Si$^{4+}$ reduction.
Si$^{4+}$ in silica can also dissolve into dense hydrogen liquid as SiH$_4$ as shown in our experiments with fayalite below (Extended Data Fig.~\ref{fig:SI:SiH4} and Supplementary Discussion~\ref{SI:section:fay-50H}) and ref.~\cite{shinozaki2014formation}.
Some silica still remained after heating, and the observation is likely due to lower-temperature heating below melting.
Raman measurements of the sample after laser heating at high pressure detected H$_2$O (Extended Data Fig.~\ref{fig:SI:map}). %
Consistent results were obtained from other runs at 14--39~GPa (Supplementary Discussion~\ref{SI:section:silica-runs}).

When fayalite, Fe-endmember olivine ($\mathrm{Fe}_2\mathrm{SiO}_4$), was heated to 3017~K at 6~GPa, it broke down completely to form cubic metal phases (fcc, bcc, and B2) (Fig.~\ref{fig:xrd}b).
The fcc phase is FeH$_x$ with volumes indicating $x$ as high as 2 to 2.3 (Extended Data Fig.~\ref{fig:SI:FeH-vol}; Supplementary Discussion~\ref{SI:section:fay-fayfield}).
In this experiment without Mg, the silicate melt is completely consumed by reaction with H and the released O reacts with H to form H$_2$O, which is confirmed by Raman spectroscopy (Supplementary Discussion~\ref{SI:section:fay-fayfield}).

Above 10~GPa, fayalite breaks down to form silica and FeH$_x$ (Supplementary Discussion~\ref{SI:section:fay-spifield} and Extended Data Fig.~\ref{fig:SI:fay21}). 
No clear evidence for an Fe-Si phase was detected.
The presence of H$_2$O is confirmed via Raman spectroscopy. %
The existence of silica diffraction lines after heating indicate that some Si remains oxidized at this higher pressure range. 
When fayalite (which has 10 times as much Fe$^{2+}$ as San Carlos olivine) is used, much more H$_2$O can be rapidly produced, likely because it is much easier for Fe$^{2+}$ to be reduced by H than Si$^{4+}$ and Mg$^{2+}$.
The rapid production of H$_2$O will raise the activity of H$_2$O and therefore suppress the reduction of Si$^{4+}$, which is more difficult than the reduction of Fe$^{2+}$. 
In runs where fayalite was melted with a much smaller amount of hydrogen (an Ar medium with 50\% H$_2$; Supplementary Discussion~\ref{SI:section:fay-50H}), SiH$_4$ was detected in Raman measurements (Extended Data Fig.~\ref{fig:SI:SiH4}). 
Si remains oxidized in SiH$_4$, and therefore water production can continue under more oxidizing conditions through the formation of SiH$_4$ (reaction~\ref{eq:ol-Si-dissolve}).

\section{SIL runs}\label{SI:section:silica-runs}

\paragraph{FeH$_x$} Although with a larger uncertainty due to the weak diffraction intensity of FeH$_x$, in (run SIL-1), the fcc phase showed a unit-cell volume of 50.7$\pm$0.2~\AA$^3$ at 14~GPa after temperature quench, slightly lower than FeH$_x$ with $x = 1$.
Similar results were also seen in other SIL runs (Extended Data Fig.~\ref{fig:SI:SiO2}).

\paragraph{Coesite/Stishovite}
Because SiO$_2$ does not couple efficiently with the near infrared laser beam but is heated mainly through the coupling of Fe metal particles nearby, the heating was less efficient than in olivine, which contains Fe and directly couples with the laser beams.
In these runs, temperature was not high enough to melt silica at this pressure (Extended Data Fig.~\ref{fig:SI:PT}).
Therefore, some silica phases remained after heating.
Some coesite peaks were identified in some runs (SIL-1 and SIL-2) despite the fact that the pressures were higher than the stable condition for coesite.
Similar observation was also made in the FAY-3 and FAY-4 runs.
This may indicate that hydrogen could affect the stable conditions for coesite.

\paragraph{Fe$_{1-y}$Si$_y$} 
In all SIL runs some Si$^{4+}$ was reduced to Si$^0$ which alloyed with Fe present to form B2~Fe$_{1-y}$Si$_y$. 
Although due to the coupling considerations mentioned above along with the high melting temperature of silica, the amount (and thus diffraction intensity) was quite small. 
Raman modes of O-H bonding were detected around the heated area but not in unheated areas, suggesting that the SiH$_4$ formation (reaction~\ref{eq:ol-Si-dissolve}) may play a major role for the H$_2$O formation.

\section{FAY-1 and 2}\label{SI:section:fay-fayfield}

\paragraph{Fe$_{1-y}$Si$_y$}
A B2 phase with a unit-cell volume 7.2\% larger than stoichiometric FeSi (or $y=0.5$, ref.~\citeS{fischer2014equations}) is observed (Extended Data Fig.~\ref{fig:SI:fay1}). 
This is likely due to a slight excess of Fe (i.e. Fe/Si$>$1) which has a higher atomic radius than Si, rather than hydrogenation of FeSi which has not been observed in previous experiments with Fe-Si alloys under a H medium \citeS{fu2022fesiH}. 

\paragraph{FeH$_x$} 
In runs FAY-1 and FAY-2, the volume of the fcc metal is 23.6--27.5\% larger than the volume of stoichiometric FeH reported at this pressure ($\sim$51.3~\AA$^3$) (ref.~\citeS{narygina2011}) after heating. 
Following the method of ref.~\protect\citeM{piet2023}, we obtained $x = 2\textrm{--}2.3$ for the FeH$_x$ (Extended Data Fig.~\ref{fig:SI:FeH-vol}).
A very large volume expansion for fcc was also found for (Fe,Ni)H$_x$ and FeH$_x$ melted at pressures above 77~GPa (ref.~\protect\citeM{piet2023}).
In that case it was estimated that $x$ was 1.8, slightly smaller but comparable to our observation.
Upon decompression to 1~bar, the fcc phase disappears, as expected for FeH$_x$, which is known to lose hydrogen below $\sim$3.5~GPa (ref.~\citeS{badding1991}) (Extended Data Fig.~\ref{fig:SI:fay1}).

\paragraph{Bcc phase}
A bcc phase with a unit-cell volume 1.7\% larger than pure bcc Fe metal was observed. 
While bcc Fe metal is not observed in the pure Fe-H system at pressures above 3.5~GPa and temperatures this high \citeS{narygina2011}, ref.~\citeS{lin2002iron} showed that a small amount of Si can stabilize bcc Fe beyond its standard stability field. 
It is likely our observation of the bcc phase is due to slight incorporation of Si. 
The volume expansion of 1.7\% may be due to hydrogenation but is far smaller than the $\sim$20--25\% volume expansion seen in FeH phases \citeS{badding1991}\textsuperscript{,} \citeS{narygina2011}, consistent with previous observations that Si (which may be stabilizing this structure) inhibits the hydrogenation of Fe metal \citeS{fu2022fesiH}.
Upon decompression to 1~bar, the B2 and bcc phases remain stable.

\section{FAY-3, 4, and 5}\label{SI:section:fay-spifield}

Above $\sim$18~GPa without H, fayalite is known to break down into FeO and stishovite \citeM{ohtani1979melting}. 
We also observed the same trend in our experiments in the presence of hydrogen.
However, in our hydrogen case, FeO reacts with hydrogen and forms FeH$_x$ (Extended Data Fig.~\ref{fig:SI:fay21}).
Somewhat lower degree of Si loss was found in these runs.  
At this high pressure range, a much smaller volume of hydrogen can be heated conductively because of the thinning of the hydrogen medium, resulting in a larger axial thermal gradient in LHDAC\@.
If this is the case, a smaller amount of liquid hydrogen may not be able to dilute H$_2$O enough to maintain a low H$_2$O activity in the heated area, reducing the extent of hydrogen-silicate reaction.

\paragraph{Silica}
Silica is present as both stishovite and coesite. 
When the molten sample is being temperature quenched, because of the positive Clapeyron slope of the coesite-to-stishovite transition, coesite may crystallize first at higher temperatures followed by crystallization of stishovite at lower temperatures \citeS{akimoto1977high}. 
Upon decompression to 1~bar, both coesite and stishovite remain. 
After decompression to 1~bar, stishovite was found to have slightly larger unit-cell volume ($V = 46.67\pm 0.08$~\AA$^{3}$) compared to 46.50~\AA$^{3}$ reported by ref.~\citeS{sinclair1978single} for anhydrous stishovite.
H$_2$O can be incorporated in the crystal structure of stishovite and increase the unit-cell volume
\citeS{nisr2020}.
A previous study \citeS{nisr2020} reported the relationship between the unit-cell parameters and water content in hydrous stishovite.
Utilizing that calibration, the change in unit-cell volumes implies $\sim$0.7~wt\% H$_2$O in stishovite. 
However, the minimal change in the axial ratio relative to anhydrous stishovite ($c/a = 0.640$ compared to $c/a = 0.641$ reported by ref.~\citeS{andrault1998pressure}) %
does not align with an increase in the axial ratio reported in ref.~\citeS{nisr2020} for hydrous stishovite. 
Therefore, the estimation of H$_2$O content should be interpreted with caution.

\paragraph{FeH$_x$} 
At this pressure range, we observed $x = 0.8\textrm{--}1$ for fcc~FeH$_x$.
Upon decompression to 1~bar, fcc~FeH$_x$ reverts to bcc~Fe metal. 
The unit-cell parameter of the bcc phase is 2.8677$\pm$0.0003~\AA, which agrees with the expected value of pure Fe metal of 2.867~\AA\ (ref.~\citeS{rotter1966ultrasonic}), suggesting that all hydrogen has left the crystal structure of Fe metal upon decompression.

\section{FAY-6, 7, and 8}\label{SI:section:fay-50H}

Additional experiments on fayalite were conducted at similar conditions to runs FAY-1, 2, 3, and 4, but with a medium of a 1:1 ratio (by volume) of H:Ar instead of pure H$_2$. 
The solid phase products were the same as in the pure H-medium experiments.
Not only does the mixed medium allow us to examine the reaction for low H$_2$ concentration, the setup also enhances the thermal and mechanical stability of the medium, particularly during decompression to obtain Raman spectra of SiH$_4$ (Extended Data Fig.~\ref{fig:SI:SiH4}).

\section{Estimation for the pressure-temperature conditions of the atmosphere-interior boundaries of super-Earths and sub-Neptunes}\label{SI:section:PT-sub-Neptune}

In order to define the planet population in which H$_2$O production takes place we examined in which planets the silicate-hydrogen boundary meet the experimental pressure-temperature range along their thermal evolution. 

We run interior evolution models of planets with rocky core and H,He envelopes, based on ref.~\cite{vazan2018Contribution}, varying the planet mass, envelope mass, and planet initial thermal state (derived from planet formation assumptions). 
The estimation shows that rocky super-Earth to sub-Neptune planets, in the mass range of 3--15\,$M_{E}$ with 2--20~wt\% of gas (H, He) are expected to experience H$_2$O production. %
In planets with $M_p < 3\,M_{E}$ and/or $<$2\% H$_2$, the core-envelope boundary (CEB) is below the experimental pressure range, while in planets with $M_p > 15\,M_{E}$ and/or $>$20\% H$_2$ temperature is above the experimental range.
The planet mass and envelope mass ranges found here are an order of magnitude estimate, and specific planet formation or structure evolution conditions can vary these values slightly.

Based on the model results, we constructed structure-evolution profiles of super-Earth and sub-Neptune planets under various conditions. 
In Fig.~\ref{fig:SI:amb-PT}, we show an example of such structure-evolution profile for a $5\,M_{E}$ super-Earth planet with 5\% H + He envelope. 
Shown is the temperature (color) profile in the interior from center to 1~bar pressure, as a function of time. 
The experimental pressure-temperature range is shown in black (dotted and solid curves, respectively) and the CEB is in dashed red. 
As can be seen, the CEB of this planet stays in water production regime for many giga years. 
Importantly, heat transport in most of the interior is found to be by large scale convection along the evolution track. 

\section{Estimated quantities of reactants and products in the LHDAC experiments}\label{SI:section:reactant-product-quantities}

The reaction observed in our experiments on olivine melt with Fe metal in a hydrogen medium can be described as the following:
\begin{equation}
\begin{split}
\mathrm{(Mg_{0.9}Fe_{0.1})_2SiO_4} + (a-0.2)\mathrm{Fe} + [2(2-z) + 0.5x( a - (z/y) )]\mathrm{H_2}  
\\
\rightarrow 2\mathrm{MgO} + (z/y)\mathrm{Fe}\mathrm{Si}_y + (a - (z/y))\mathrm{FeH}_x + (1-z)\mathrm{SiH_4} + 2\mathrm{H_2O},\label{eq:complete-reaction}
\end{split}
\end{equation}
where $a$ is the total amount of Fe in the system including Fe metal loaded together with olivine and Fe$^{2+}$ in olivine, which is 0.86.
$y$ and $x$ are the amounts of Si and H in Fe$_{1-y}$Si$_y$ and FeH$_x$ after reaction, respectively.
They can be obtained from the measured unit-cell volumes combined with composition-volume relations \protect\citeM{piet2023}\textsuperscript{,}\citeS{fischer2014equations}. 
$z$ is the fraction of Si reduced to metal and can be estimated by comparing diffraction intensities and amount of Si in Fe-Si alloy, i.e., $y$.

In an LHDAC experiment, amount of heated medium, H$_2$ in our case, is difficult to measure.
However, some observational constraints can be used for an estimation.  
The volume of warm dense liquid hydrogen involved in the reaction should be limited to that in the grain boundaries of the heated part of the olivine + Fe metal sample foil and the medium layers adjacent to the heated sample surfaces along the loading axis of DAC (Fig.~\ref{fig:fib}a).  
It should be limited laterally as well, to the laser beamsize, because of the radial thermal gradients in LHDAC\@.  
From these constraints, we found that 4.5--5.7~wt\% H$_2$ existed in the heated volume together with $\sim$76~wt\% silicate and $\sim$19~wt\% Fe metal in our sample setup before the reaction.

The amount of H$_2$O produced from the reaction can be much better estimated.
First, concerning the Si effect only, both Fe-Si alloy formation and SiH$_4$ formation produce the same amount of H$_2$O per olivine as these reactions release the same amount of O per olivine from molten silicate to liquid hydrogen.
Second, in our olivine + Fe metal runs, all Si in silicate was released at lower pressures (6--8~GPa) and almost entirely at higher pressures ($>$25~GPa).
Therefore, the involved parameters, such as $x$, $y$, and $z$, in reaction~\ref{eq:complete-reaction} do not affect the estimation for the amount of total H$_2$O produced, but it depends only on the amount of melted olivine: 2 moles of H$_2$O can be produced from 1 mole of molten olivine.  
According to our estimation, 18.1(5)~wt\% H$_2$O was produced by \ref{eq:complete-reaction}.  
This calculation is provided in Supplementary Code~1.

We also provide Supplementary Code~2 for the case of partial reaction and Supplementary Code~3 for the effects of Mg/Si ratio.

\section{Segregation of iron alloys in a magma ocean}\label{SI:section:Fe-segregation}

Once the liquid Fe metal alloy is fully sequestered from the silicate melt by density to form a stratified metallic layer below the silicate melt layer, the reduction of Si may stop.
However, the formation of SiH$_4$ can continuously liberate oxygen from silicates to form water without the need to reduce Si$^{4+}$.   
On the other hand, liquid droplets of Fe metal alloy---if sufficiently small in size---could be suspended in silicate melt, delaying metal-rock segregation in the molten interior of sub-Neptunes\citeS{lichtenberg2021Redox,young2024Phase}, an effect that could be further intensified by the decrease in metal droplet density due to alloying of Fe with H and Si present in the magma ocean.
In this case, the Fe-Si alloy formation may continue to contribute to the endogenic production of water. 

\section{Endogenic water production contributed by magnesium involved reactions}\label{SI:section:Mg-reactions}

In ref.~\cite{horn2023reaction} and this study, MgO remained after the hydrogen-magma reaction, mainly because the temperatures were lower than the melting of the MgO component.
In more recent experiments where MgO and Fe metal were melted together in a hydrogen medium, it was found that the Mg--O bond breaks down to form Mg hydrides \cite{kim2023Stability}.
The reaction also results in the release of O and formation of water, similar to the case of SiH$_4$ formation in this study.
However, because much less Mg hydride was produced in that reaction, its contribution to the total amount of endogenic water should be much smaller than that from Si release.
On the other hand, because Mg hydride formation occurs at higher temperatures, endogenic water production can occur from at least $\sim$5000~K\@.

The study \cite{kim2023Stability} also observed crystallization of Mg$_2$FeH$_6$ after heating at pressures below 13~GPa.
Although the melting temperature of Mg$_2$FeH$_6$ is not known to our knowledge, from the melting temperatures of related materials, such as MgH$_2$ and FeH (refs~\citeS{moser2011Pressure,fukai2003Phase}) at the pressure range, it should be lower than 1500~K if the material undergoes incongruent melting.
Therefore, Mg$_2$FeH$_6$ is unlikely to impact the hydrogen-magma reaction we report here until a sub-Neptune cools down to sufficiently low temperatures to crystallize the phase.
On the other hand, the hydride bond formation will lower the melting temperature of the magma and extend the molten state of the interior and facilitating more efficient mixing and chemical reacting between the hydrogen and magma.

\section{Impact of SiH$_4$ and MgH$_2$ on the dynamics of the interior}\label{SI:section:SiH4-dynamics}

Although, to our knowledge, no direct experimental measurements or computational simulations have been performed, experiments have shown that SiH$_4$ and H$_2$ are miscible in the fluid state up to 6~GPa and 300~K (ref.~\citeS{wang2009High}).
Furthermore, SiH$_4$ and H$_2$ can form a solid solution (i.e., miscibility in the solid state) at high pressures\citeS{wang2009High,strobel2009Novel}.
Therefore, under the $P$--$T$ conditions relevant to the CEB of sub-Neptunes, SiH$_4$ and H$_2$ are expected to exist as a single fluid.

To our knowledge, the density of fluid SiH$_4$ is not known at the $P$--$T$ conditions of the CEB\@.
However, useful insights can be gained by comparing the solid phases of H$_2$O and SiH$_4$, as their equations of state have been measured at high pressures and 300~K (refs~\citeS{sugimura2008Compression,degtyareva2007Crystal}).
We find that the density of SiH$_4$ is 20--30\% lower than that of H$_2$O over the pressure range considered in this study.
Similarly, the density of MgH$_2$ is slightly lower than that of H$_2$O (refs~\citeS{moriwaki2006Structural,cui2008Structural}).
This is primarily due to the significantly larger molar volumes of SiH$_4$ and MgH$_2$ compared with H$_2$O.

There are several explanations for the larger molar volumes of these hydrides relative to H$_2$O.
The most notable is the difference in the oxidation state of hydrogen: in SiH$_4$ and MgH$_2$, hydrogen exists as H$^{-}$, whereas in H$_2$O, it is H$^{+}$.
Because H$^{-}$ has two electrons and only one positive nuclear charge, it has a much larger volume (ionic radius of 1.34~\AA) than H$^{+}$ (ionic radius of 0.18~\AA), resulting in significantly larger volumes for SiH$_4$ and MgH$_2$ compared to H$_2$O.

The larger molar volumes of MgH$_2$ and SiH$_4$ result in lower densities relative to H$_2$O under the relevant $P$–$T$ conditions. 
As a result, a mixture containing MgH$_2$, SiH$_4$, and H$_2$O has a lower density than pure H$_2$O. 
In planetary interiors, large-scale convection is an efficient mechanism for heat and material transport. 
This process is driven by buoyancy forces arising from density gradients due to thermal or compositional differences \citeS{kippenhahn1991Stellar}. 
Since the strength of convective-mixing is inversely related to the density variation between the fluid and its surroundings (e.g., Appendix in ref.~\citeM{vazan2015convection}), the reduced density of the mixture enhances the efficiency of convective-mixing and prolongs its activity.

If, under certain conditions the MgH$_2$, SiH$_4$, and H$_2$O mixture would result in higher density than that of a pure water that we considered in our simulations, a semi-convection (double diffusive convection \citeS{turner64DDC}) may develop instead of large-scale convection. 
Semi-convection would limit the mixing efficiency and consequently the water production may stop earlier.
However, these conditions were not found in our examination.

\clearpage

\bibliographystyleS{naturemag-doi}
\bibliographyS{dis.bib,add.bib}

\newpage
\renewcommand{\figurename}{\bf Fig.}
\renewcommand{\tablename}{\bf Table}

\end{document}